\documentclass[
amsmath, %
floatfix, %
twocolumn, %
reprint, %
prl, %
aps, %
citeautoscript, %
]{revtex4-2} 
\renewcommand{\thispagestyle}[1]{}
\usepackage[]{newtxtext} 
\usepackage[subscriptcorrection,nosymbolsc,smallerops,bigdelims]{newtxmath} 
\DeclareMathAlphabet{\mathcal}{OMS}{cmsy}{m}{n} 
\DeclareMathAlphabet{\mathbcal}{OMS}{cmsy}{b}{n} 
\usepackage{bm}

\usepackage[utf8]{inputenc}
\usepackage[T1]{fontenc}

\usepackage[]{graphicx}
\usepackage{latexsym}
\usepackage{color}
\usepackage{mathtools}
\usepackage[section]{placeins}
\usepackage[]{xcolor}
\usepackage{bm}
\usepackage{bbold}
\usepackage{multirow}
\usepackage{siunitx}
\usepackage{hyperref}
\hypersetup{
	colorlinks,
	linkcolor={blue!90!black},
	citecolor={blue!90!black},
	urlcolor=	{blue!90!black}
}
\let\oldeqref\eqref
\renewcommand*{\eqref}[1]{%
	\hyperref[#1]{\oldeqref{#1}}%
}

\usepackage{floatrow}
\usepackage{placeins}

\newcommand{\mr}[1]{\mathrm{#1}}

\mathchardef\mhyphen="2D

\DeclarePairedDelimiter\abs{\lvert}{\rvert}

\DeclarePairedDelimiterX{\comm}[2]{\lbrack}{\rbrack}{#1, #2}
\DeclarePairedDelimiter\ket{\lvert}{\rangle}

\DeclarePairedDelimiterX{\braket}[2]{\langle}{\rangle}{#1\delimsize\vert #2}
\DeclarePairedDelimiterX{\ketbra}[2]{\rvert}{\lvert}{#1 \delimsize\rangle\!\delimsize\langle #2}
\DeclarePairedDelimiterX{\matrixel}[3]{\langle}{\rangle}{#1 \delimsize\vert #2 \delimsize\vert #3}

\newcommand{\hc}{\mr{H.c.}}

\DeclareMathOperator\erf{erf}

\newcommand{\eqnref}[1]{Eq.~\eqref{#1}}

\newcommand{\subfigref}[2]{Fig.~\hyperref[#1]{\ref*{#1}(#2)}}
\newcommand{\isubfigref}[2]{Figure~\hyperref[#1]{\ref*{#1}(#2)}}
\newcommand{\subfigsref}[3]{Figs.~\hyperref[#1]{\ref*{#1}(#2)}-\hyperref[#1]{\ref*{fig:#1}(#3)}}

\usepackage[low-sup]{subdepth}
\usepackage{ragged2e}

\definecolor{cbred}{HTML}{e31a1c}
\definecolor{cbgreen}{HTML}{33a02c}
\definecolor{cbblue}{HTML}{176aa7}
\definecolor{cborange}{HTML}{ff7f00}
\definecolor{cbviolet}{HTML}{6a3d9a}
\definecolor{cbbrown}{HTML}{b15928}

\definecolor{cblred}{HTML}{fb9a99}
\definecolor{cblgreen}{HTML}{b2df8a}
\definecolor{cblblue}{HTML}{a6cee3}
\definecolor{cblorange}{HTML}{fdbf6f}
\definecolor{cblviolet}{HTML}{cab2d6}
\definecolor{cblbrown}{HTML}{ffff99}

\usepackage{tikz}
\usetikzlibrary{quantikz2}

\usepackage[activate={true,nocompatibility},final,tracking=alltext,kerning=true,spacing=true,protrusion=true,factor=1100,stretch=5,shrink=9,selected=true ,letterspace=-0]{microtype}
\usepackage{orcidlink}

\definecolor{CBRed}{RGB}{228,26,28}
\definecolor{CBBlue}{RGB}{55,126,184}
\definecolor{CBGreen}{RGB}{77,175,74}
\definecolor{CBPurple}{RGB}{152,78,163}
\definecolor{CBOrange}{RGB}{255,127,0}
\definecolor{CBYellow}{RGB}{255,255,51}
\definecolor{CBBrown}{RGB}{166,86,40}
\definecolor{CBPink}{RGB}{247,129,191}
\definecolor{CBGrey}{RGB}{153,153,153}

\newcommand{\symRedCircle}{\textcolor{CBRed}{\ensuremath{\bullet}}}
\newcommand{\symBlueSquare}{\textcolor{CBBlue}{\ensuremath{\blacksquare}}}
\newcommand{\symGreenTri}{\textcolor{CBGreen}{\ensuremath{\blacktriangle}}}
\newcommand{\symOrangeStar}{\textcolor{CBOrange}{\ensuremath{\star}}}
\newcommand{\symPurpleDiam}{\textcolor{CBPurple}{\ensuremath{\blacklozenge}}}

\newcommand*{\AcoustAmpl}{A}

\begin{document}

\title{Hybrid acousto-optical spin control in quantum dots}

\author{Mateusz Kuniej\,\orcidlink{0000-0001-5476-4856}}
\affiliation{Institute of Theoretical Physics, Wroc\l{}aw University of Science and Technology, 50-370 Wroc\l{}aw, Poland}

\author{Pawe{\l} Machnikowski\,\orcidlink{0000-0003-0349-1725}}
\affiliation{Institute of Theoretical Physics, Wroc\l{}aw University of Science and Technology, 50-370 Wroc\l{}aw, Poland}

\author{Micha{\l} Gawe{\l}czyk\,\orcidlink{0000-0003-2299-140X}}
\affiliation{Institute of Theoretical Physics, Wroc\l{}aw University of Science and Technology, 50-370 Wroc\l{}aw, Poland}

\begin{abstract}
    Mechanical degrees of freedom very weakly couple to spins in semiconductors. The inefficient coupling between phonons and single electron spins in semiconductor quantum dots (QDs) hinders their integration into on-chip acoustically coupled quantum hybrid systems. We propose a hybrid acousto-optical spin control method that circumvents this problem and effectively introduces acoustic spin rotation to QDs, complementing their rich couplings with external fields and quantum registers.
    We show that combining continuous-wave detuned optical coupling to a trion state and acoustic modulation results in spin rotation around an axis defined by the acoustic field. The optical field breaks spin conservation, allowing phonons to drive transitions between disrupted spin states when at resonance with the Zeeman frequency.
    Our method is compatible with pulse sequences that mitigate quasi-static noise effects, which makes trion recombination the primary limitation to gate fidelity under cooled nuclear-spin conditions. Numerical simulations indicate that spin rotation fidelity can be very high, if the trion lifetime is long and Zeeman splitting is sufficiently large, with a currently feasible 50~ns lifetime and 44~GHz splitting giving 99.9\% fidelity.
    Applying our advancement could enable acoustic QD spin state transfer to diverse solid-state systems and transduction between acoustic, optical, and microwave domains, all within an on-chip integration-ready setting.
\end{abstract}

\maketitle

{\it Introduction---}Acoustic waves have long been used for signal filtering, mixing, and duplexing in classical electronics like radio-frequency (RF) communication systems. These components are foundational in modern mobile and wireless technologies, where coupling mechanical waves to electrical signals enables highly selective signal processing in compact architectures \cite{Campbell1989, Hashimoto2000}. This widespread application in classical RF systems is accompanied by advancements in coherent acoustic generation \cite{Delsing2019} with increasing frequency limits \cite{Choquer2022, Zheng2020, Zhou2023}. This is followed by explorations at the quantum level, where mechanical degrees of freedom can couple to single charges and spins in low-dimensional systems \cite{Priya2023}. Among such systems, epitaxial semiconductor quantum dots (QDs) offer a highly controllable environment for investigating acoustic control of charge dynamics \cite{Schulein2013} and optical effects, including acoustic control of single-photon scattering \cite{Weiss2021} and emission~\cite{Buhler2022}.

The next logical step is to extend acoustic control to spin states. Spins in QDs present a substantial advantage for quantum information processing due to relatively long coherence times. While defect centers offer a well-established acoustic interface to spin states \cite{Maity2020}, QDs stand out as solid-state quantum emitters for their brightness and purity of single photons \cite{Lodahl2015,Somaschi2016,Tomm2021}. Spins in QDs assure compatibility with photonic interfaces \cite{Yilmaz2010} in the optical range and show potential for generating entanglement with flying qubits \cite{Delley2017} and producing photonic cluster states \cite{Lindner2009,Economou2010,Michaels2021,Cogan2023}. Good quality QDs can couple confined spins to homogeneous nuclear spin ensembles, allowing quantum information transfer to nuclear degrees of freedom \cite{Chekhovich2020,Shofer2025,Appel2025}. With optical emission tunable to the telecom band on various material platforms \cite{Holewa2024, Holewa2022, Deutsch2023, Michl2023}, QDs hosting single spins seem to be the preferred physical platform for quantum networking applications. 

A challenge in achieving spin-acoustic coupling with QDs is the relatively weak intrinsic coupling between mechanical deformation and spin in semiconductors. Recent advances in integrating coherent laser sources on various chip platforms \cite{Li2022, VanGasse2019} indicate that hybrid optomechanical interaction is a viable, technologically feasible path, circumventing this obstacle. Pursuing this idea paves the way for universal acoustic transducers \cite{Schuetz2015} using spin qubits in QDs that might couple photonic, charge, nuclear, and mechanical degrees of freedom at the quantum level. Furthermore, it would provide a quantum data bus to various qubits, including superconducting circuits, one of the workhorses of today's practical quantum computing, thus opening new frontiers in scalable quantum information processing.

Here, we take an important step towards quantum coupling of multiple degrees of freedom: while potential implementations across various solid-state systems have been proposed \cite{Schuetz2015}, we address the persistent gap for single spins by proposing an acousto-optical method for single-carrier spin control. It relies on detuned optical excitation that breaks spin conservation by coupling both spin states to a trion state, and acoustic modulation resonant with the spin splitting, which can then induce spin transitions. This hybrid method enables any spin rotation using acoustic pulses during detuned continuous-wave optical coupling. We show that already a na\"ive two-pulse implementation of the Pauli-X gate can be fast enough to keep the infidelity due to the Overhauser field below 0.1\% in QDs with cooled nuclear spins \cite{SunPRL2012,EthierMajcherPRL2017,JacksonPRX2022,NguyenPRL2023}, and that pulse sequences that mitigate this quasi-static noise to the leading order can be employed. The remaining limiting factor for spin-rotation fidelity is then trion recombination, which is largely suppressed by the tiny trion occupation during operation. Thus, fast evolution, assured by a high acoustic frequency, and slow trion recombination are essential factors. We predict the recombination-caused infidelity $<0.1\%$ with the currently available acoustic frequency for a double QD (DQD) system, where a spatially indirect long-lived trion state can be used.

\begin{figure}[tb!]
    \centering
    \includegraphics[width=\columnwidth]{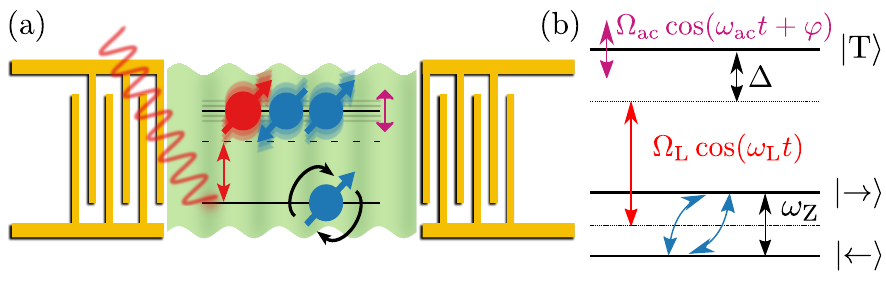}
    \caption{Idea of the acousto-optical spin rotation in a quantum dot. (a) Principle of the method: Detuned laser couples spins states to a trion state (red arrow), which is acoustically modulated (pink arrow) with a frequency equal to spin splitting, e.g., via an interdigital transducer (gold). Due to optical coupling, spin conservation is lifted, allowing for acoustic spin rotation. (b) Schematic energy structure of the three-level system with external fields. Spin states $\ket{\leftarrow}$ and $\ket{\rightarrow}$ are split in a magnetic field in Voigt configuration and optically coupled (red arrow) to the trion level $\ket{\mathrm{T}}$ with detuning $\Delta$. Acoustic modulation of the trion state is marked with the pink arrow. Blue arrows show the effective coupling between spin states.}
    \label{fig:system}
\end{figure}
    
    {\it System and model---}\isubfigref{fig:system}{a} pictures the idea outlined here, and \subfigref{fig:system}{b} presents a more detailed level diagram. The resident electron spin in a QD is subject to an external magnetic field in the Voigt configuration and off-resonantly coupled to a charged exciton (trion) state with a continuous-wave laser (red arrows). The laser frequency $\omega_{\mathrm{L}}$ is detuned from the trion energy $\hbar\omega_{\mathrm{T}}$ by $\Delta = \omega_{\mathrm{L}}-\omega_{\mathrm{T}}$. According to selection rules, the Voigt configuration and $\sigma_{+}$-polarized laser ensure equal coupling of both $\ket{\rightarrow}$ and $\ket{\leftarrow}$ states to the same trion state $\ket{\mathrm{T}}$. This coupling aims to break spin conservation by coupling both spin states to a common state, thereby slightly mixing them (such $\Lambda$-systems utilize the third state for transitions within the doublet \cite{Scully1997,GolterPRX2016}). Detuning prevents noticeable trion occupation, which could be detrimental due to the finite trion recombination time $\tau$, causing dephasing. The optical transition (trion) energy is modulated by a coherent monochromatic acoustic wave via deformation potential, like a surface acoustic wave (SAW) generated by an interdigital transducer (IDT). As we show in the following, tuning the acoustic energy $\hbar\omega_{\text{ac}}$ to resonance with the spin splitting $\hbar\omega_{\mathrm{Z}}$ induces spin rotation.
    
    The Hamiltonian of the system is given by
    \begin{equation}
        H(t) = H_0 + H_{\mathrm{ac}}(t) + H_{\mathrm{L}}(t) + H_{\delta}
            \equiv H_{\mathrm{c}}(t) + H_{\delta} ,
    \end{equation}
    where $H_0 = \hbar\omega_{\mathrm{Z}}\ketbra{\rightarrow}{\rightarrow} +\hbar\omega_{\mathrm{T}}\ketbra{\mathrm{T}}{\mathrm{T}}$ describes the system's free evolution. Further, $H_{\mathrm{ac}}$ includes acoustic modulation. With acoustic wavelengths exceeding the QD size and corresponding piezoelectric fields typically screened by a metallic layer between the QD membrane and the piezoelectric material \cite{Nysten2021PhD}, we deal with mechanical, spatially homogeneous, quasi-adiabatic band shifts without significant wavefunction distortion or level hybridization. All orbital states are modulated, but since the trion is modulated with a different strength than single electron states, $H_{\mathrm{ac}}$ can be written as pure trion modulation
    \begin{equation}
        H_{\mathrm{ac}}(t) = - \hbar\AcoustAmpl_{\mathrm{ac}}f(t)\cos\left(\omega_{\mathrm{ac}}t - \varphi\right)\ketbra{\mathrm{T}}{\mathrm{T}},
    \end{equation}
    where $f(t)$ is an acoustic pulse envelope with amplitude $\AcoustAmpl_{\mathrm{ac}}$ (See End Matter Sec.~A). $H_{\mathrm{L}}(t) = -\bm{d}\cdot\bm{E}(t)$ describes the optical coupling in the dipole approximation, where $\bm{d}$ is the dipole moment operator, and $\bm{E}(t) = \bm{E}_0(t)\cos(\omega_{\mathrm{L}}t)$ is the laser electric field. Those three terms form our control Hamiltonian $H_{\mathrm{c}}(t)$. The last term, $ H_{\delta} = (\hbar/2)\, \bm{\delta}\cdot\bm{\sigma}$ with $\bm{\sigma}$ denoting the vector of Pauli matrices in the basis $\{\ket{\rightarrow},\ket{\leftarrow} \}$, describes the effect of quasi-static Overhauser noise $\bm{\delta}$ from the isotropic nuclear spin environment \cite{Merkulov2002}, assumed to have independent zero-mean normal distributions with standard deviation $\sigma_\delta$ per component.  
    In the rotating frame (with respect to $\omega_{\mathrm{Z}}$ and $\omega_{\mathrm{L}}$) and rotating wave approximation, the Hamiltonian can be written as \cite{Scully1997}
    \begin{subequations}\label{eq:Hfull}
        \begin{align}
            \!\!\!H_{\mathrm{c}}(t) ={}& -\hbar\Delta\ketbra{\mathrm{T}}{\mathrm{T}} + H_{\mathrm{ac}}(t) \notag \\ &+ \frac{\hbar}{2} A_{\mathrm{L}} \left(e^{i\omega_{\mathrm{Z}}t}\ketbra{\rightarrow}{T} + \ketbra{\leftarrow}{T} + \hc\right),\\
            \!\!\!H_{\delta}(t) ={}& \frac{\hbar}{2}\,\delta_z\sigma_z + \frac{\hbar}{2}\left[\left( \delta_x + i\delta_y \right) e^{i\omega_{\mathrm{Z}}t}\ketbra{\rightarrow}{\leftarrow}+\hc\right],
        \end{align}
    \end{subequations}
    where $A_{\mathrm{L}}$ is the laser amplitude.
To find the dissipative evolution of the system with trion recombination, we numerically solve the Lindblad master equation for the density matrix $\rho$
\begin{equation}\label{eq:lindblad}
    \Dot{\rho}(t) = \frac{1}{i\hbar}\left[H(t), \rho(t)\right] + \frac{1}{\tau} \sum_{i} \left(L_{i}\rho(t)L^{\dagger}_{i} - \frac{1}{2}\left\{L^{\dagger}_{i}L_{i}, \rho(t)\right\} \right),  
\end{equation}
with Hamiltonian \eqref{eq:Hfull} and the initial state $\rho(t_0) = \ketbra{\rightarrow}{\rightarrow}$, where $L_{i} = \ketbra{i}{\mathrm{T}}$ are the jump operators for $i = \rightarrow$, $\leftarrow$.

\begin{figure*}[tb!]
    \centering
    \includegraphics[width=0.9\textwidth]{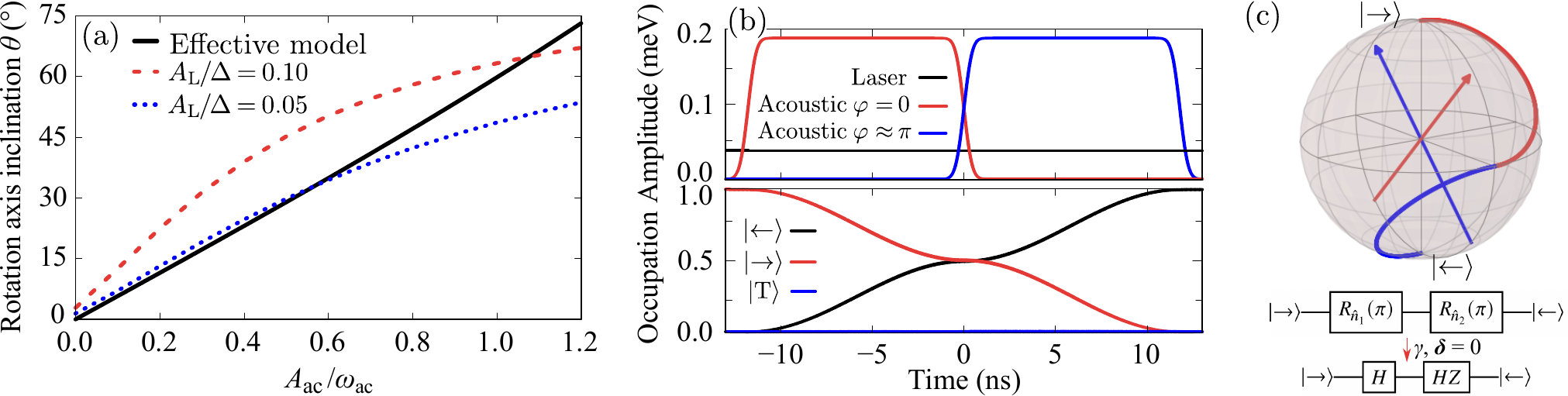}
    \caption{Spin state control. Acousto-optical driving rotates the state around an inclined axis. (a) Comparison of the inclination angle of the spin rotation axis between the effective model (solid line) and the full numerical solution for two different laser amplitudes (dashed and dotted lines). All inclinations up to 45 degrees are conveniently achievable. (b) Na\"ive implementation of the X gate. Evolution of spin states (bottom panel) under a continuous-wave detuned laser and two acoustic pulses of close to opposite phases (shown in the top panel) tuned to provide $\pi$-rotations around 45-degree inclined axes. Parameters: $\hbar\omega_{\mathrm{Z}} = 0.182$~meV, $\hbar\Delta \approx 0.766$~meV, $\hbar A_{\mathrm{L}} \approx 37.9$~\textmu{}eV, $\hbar A_{\mathrm{ac}} \approx 0.189$~meV, acoustic pulse duration $\approx11.8$~ns, and $\hbar\omega_{\mathrm{ac}} = 0.182$~meV; dissipation and noise neglected (c) Evolution on the Bloch sphere. Arrows mark rotation axes related to the first ($\varphi=0$; red) and second ($\varphi\approx\pi$; blue) acoustic pulse. At the bottom, explicit operations and an equivalent quantum circuit are shown.}
    \label{fig:protocol}
\end{figure*}

{\it Effective Hamiltonian---}To reveal how the laser and modulation jointly drive the spin evolution, we perform a unitary transformation, $\widetilde{H}(t) = e^{iS(t)}H(t)e^{-iS(t)} - \hbar\Dot{S}(t)$, with respect to $H_{\mathrm{ac}}(t)$,
\begin{equation}
    \begin{split}
        S(t) {}&{}= 
        \frac{1}{\hbar}\int_{0}^{t}\mathrm{d}t'H_{\mathrm{ac}}(t') \approx  -\AcoustAmpl(t)\sin(\omega_{\mathrm{ac}}t - \varphi)\ketbra{\mathrm{T}}{\mathrm{T}},
    \end{split}
\end{equation}
where $\AcoustAmpl(t) = \AcoustAmpl_{\mathrm{ac}}f(t)/\omega_{\mathrm{ac}}$, and we assumed slowly-varying $f(t)$. We get $\widetilde{H}_{\delta}(t)=H_{\delta}(t)$, and, using the Jacobi-Anger formula $e^{iA\sin(x)}=\sum_{n}J_n(A)e^{inx}$,
\begin{multline}
        \!\!\!\!\!\widetilde{H}_{\mathrm{c}}(t) = -\hbar\Delta\ketbra{\mathrm{T}}{\mathrm{T}} + \frac{\hbar}{2} A_{\mathrm{L}}\sum_{n=-\infty}^{+\infty} \Big[ J_n\left(\AcoustAmpl(t)\right)  e^{in(\varphi+\pi)} \\ 
        \times \left(e^{i(\omega_{\mathrm{Z}} - n\omega_{\mathrm{ac}})t}\ketbra{\rightarrow}{\mathrm{T}} + e^{-in\omega_{\mathrm{ac}}t}\ketbra{\leftarrow}{\mathrm{T}} \right) + \hc \Big],
    \label{eq:hamiltonianJacobiAnger}
\end{multline}
where $J_n$ are the Bessel functions of the first kind. For $\omega_{\mathrm{Z}} = n\omega_{\mathrm{ac}}$ with integer $n$, corresponding to $n$-phonon processes, we get a secular (nonoscillating) driving term that dominates the evolution. The strongest coupling corresponds to $\omega_{\mathrm{Z}} = \omega_{\mathrm{ac}}$, which we discuss further.

In the secular approximation, and during the acoustic pulse plateau, $\AcoustAmpl(t)=\AcoustAmpl$, we obtain a time-independent control Hamiltonian
\begin{multline}
        \!\!\!\!\widetilde{H}_{\mathrm{c}} = -\hbar\Delta\ketbra{\mathrm{T}}{\mathrm{T}} \\
        + \frac{\hbar}{2} A_{\mathrm{L}}\left(J_1(\AcoustAmpl)e^{i(\varphi+\pi)}\ketbra{\rightarrow}{\mathrm{T}} + J_0(\AcoustAmpl)\ketbra{\leftarrow}{\mathrm{T}} + \hc\right),
    \label{eq:secularHamiltonian}
\end{multline}
to which we use adiabatic elimination (see End Matter Sec.~B) to eliminate the weakly occupied trion state and arrive at an effective spin control Hamiltonian,
\begin{equation}\label{eq:Hspin}
    \begin{split}
        H_{\mathrm{spin}} ={}& -\frac{\hbar A^2_{\mathrm{L}}}{4\Delta} \Big\lbrace
         \left[J_0^2(\AcoustAmpl) - J_1^2(\AcoustAmpl)\right]\sigma_{\mathrm{z}} \\
         &{} + 2J_0(\AcoustAmpl)J_1(\AcoustAmpl)\left(\cos\varphi\,\sigma_{\mathrm{x}} + \sin\varphi\,\sigma_{\mathrm{y}}\right) \Big\rbrace,
    \end{split}
\end{equation}
with typical longitudinal (detuning) noise $\widetilde{H}_{\delta} = \hbar\,\delta_z\sigma_z$ \cite{Merkulov2002}. $H_{\mathrm{spin}}$ describes a Stark shift and off-diagonal coupling, causing spin rotation around an axis inclined at an angle
\begin{equation}
    \theta = \tan^{-1}\left(\frac{2J_0(\AcoustAmpl)J_1(\AcoustAmpl)}{J_0^2(\AcoustAmpl) - J_1^2(\AcoustAmpl)}\right) \stackrel{\AcoustAmpl\lesssim1}{\approx} \AcoustAmpl,
\end{equation}
depending on $\AcoustAmpl$ only, with the azimuth set by the phase~$\varphi$, at a rate $\Omega=A^2_{\mathrm{L}}[J_0^2(\AcoustAmpl) + J_1^2(\AcoustAmpl)]/(2\Delta)$.

We show the dependence of the inclination angle on $\AcoustAmpl$ in \subfigref{fig:protocol}{a}. The effective model prediction (solid black) agrees well with numerical solutions of Eq.~\eqref{eq:lindblad} (dashed and dotted lines; the axis extraction is described in the End Matter Sec.~C), indicating that the simple Hamiltonian captures the primary process in the evolution. The visible laser amplitude dependence not predicted by the effective model is due to non-secular terms that cannot be neglected for stronger optical fields. Nonetheless, acoustic field parameters fully control the spin rotation axis.

{\it Numerical results---}A fully horizontal rotation axis cannot be achieved without significant trion occupation. Still, the available rotations allow for full qubit control. As a proof of concept, we construct the $X$ gate by using two flat-top acoustic pulses shown in the top panel of \subfigref{fig:protocol}{b}. The first takes the state to the Bloch sphere's equator, while the second further to the opposite pole [\subfigref{fig:protocol}{c}]. Both pulses last $\approx11.8$~ns with switching time $\kappa=500$~ps, have amplitude $A_{\mathrm{ac}}\approx 189$~\textmu{}eV and differ only in phase by $\approx\pi$, providing $\pi$ rotations around axes inclined by $\theta=\pi/4$ but with opposite azimuths. A deviation from exactly opposite phases compensates for the short phase accumulation during the acoustic field switching. This sequence performs a full spin rotation, shown in \subfigref{fig:protocol}{b} (bottom) and \subfigref{fig:protocol}{c}. The rotation is exact when no noise ($\bm{\delta}=0$) and dissipation [$\tau\to\infty$ in Eq.~\eqref{eq:lindblad}] are considered.

Formally, we begin in the $\ket{\rightarrow}$ state, and perform rotations $R_{\hat{n}_2}(\pi)\,R_{\hat{n}_1}(\pi)$ with axes $\hat{n}_{1(2)}= (\pm\hat{x} + \hat{z})/\sqrt{2}$, as shown in \subfigref{fig:protocol}{c}. When performed coherently, it is equivalent to applying $H$ and $Z$ followed by $H$ quantum gates, giving $X$ in total. This evolution can be stopped at any time, covering all horizontal-axis rotations. Combined with the phase gate via a detuned laser and AC Stark shift, this enables complete qubit control.

\begin{figure*}[tb!]
    \centering
    \includegraphics[width=\columnwidth]{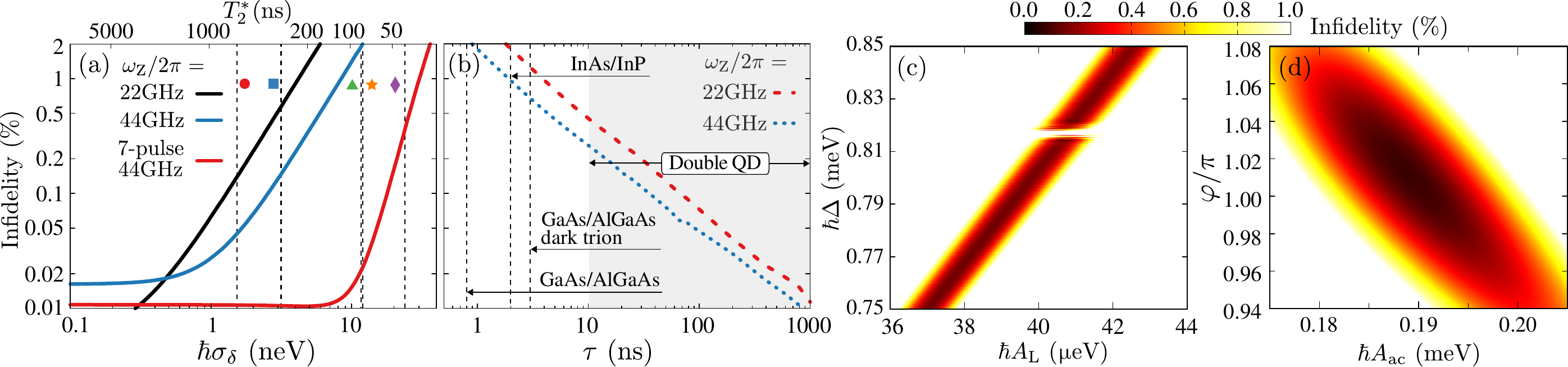}
    \caption{Fidelity and robustness of spin rotation. (a) Average infidelity of the $X$ gate due to longitudinal Overhauser noise for the na\"ive two-pulse implementation with two $\omega_{\mathrm{Z}}$ values (blue, black) and for the seven-pulse noise-mitigating sequence (red) as a function of the noise root mean square energy $\hbar\sigma_{\delta}$ (bottom axis) or corresponding spin $T_2^{*}$ time (top axis). Vertical lines mark $T_2^{*}$ for relevant systems: \symRedCircle{} Ref.~\cite{NguyenPRL2023}, \symBlueSquare{} Ref.~\cite{JacksonPRX2022}, \symGreenTri{} Ref.~\cite{SunPRL2012}, \symOrangeStar{} Ref.~\cite{NguyenPRL2023}, \symPurpleDiam{} Ref.~\cite{EthierMajcherPRL2017}. (b) Infidelity due to trion recombination as a function of the trion lifetime $\tau$ for two Zeeman splittings (dashed and dotted lines). Vertical lines mark typical $\tau$ values for relevant systems. (c) and (d) show the gate robustness to field fluctuations for fixed $\omega_{\mathrm{Z}} = 44$~GHz, $\tau = 50$~ns. (c) Infidelity versus laser amplitude and detuning for a given acoustic pulse, and (d) as a function of acoustic field amplitude and phase of the second acoustic pulse for fixed optical field parameters.}
    \label{fig:fidelity}
\end{figure*}

{\it Fidelity and robustness---}Spin rotation fidelity is limited by quasi-static noise and finite trion lifetime. In both cases, the operation time is crucial, mainly set by $\omega_{\mathrm{ac}}=\omega_{\mathrm{Z}}$, and we consider the state-of-the-art IDT-generated SAW frequencies of 22 and 44~GHz~\cite{Choquer2022, Zheng2020, Zhou2023}. Favorable impedance matching conditions enable almost lossless SAW transfer from piezoelectric LiNbO$_3$ to bonded semiconductor membranes with QDs, with frequencies increasing as the membrane thickness decreases \cite{Nysten2017,Nysten2021PhD}. Moreover, current IDTs on GaAs reach ${\sim}18$~GHz \cite{Kukushkin2006}, while optical excitation demonstrates ${\sim}30$~GHz SAWs, ${\sim}42$~GHz coherent LA phonons \cite{Ken2025}, and ${>}20$~GHz guided modes \cite{Yaremkevich2021} in GaAs, indicating the availability of tens-of-GHz acoustics in semiconductors. While a few-meV modulation amplitudes are achievable \cite{Pustiowski2015}, our scheme requires ${\sim}100$--200~\textmu{}eV, leaving room for imperfect efficiency.

We examine each decoherence source in turn. \isubfigref{fig:fidelity}{a} shows the average infidelity, $\overline{r}=1-F$, of our na\"ive two-pulse $X$ gate realized for the two acoustic frequencies as a function of the longitudinal noise strength $\hbar\sigma_{\delta}=\hbar\sqrt{2}/T_2^{*}$, where $T_2^{*}$ is the spin transverse dephasing time. $\overline{r}$ is calculated by numerically solving Eq.~\eqref{eq:lindblad} with varying $\delta_z$ while neglecting recombination, and averaging over the distribution of $\delta_z$.
Vertical lines indicate $\sigma_{\delta}$ values for self-assembled InAs/GaAs \cite{SunPRL2012,EthierMajcherPRL2017,JacksonPRX2022} and droplet-etched GaAs/AlGaAs QDs \cite{NguyenPRL2023} with cooled nuclear environments (see End Matter Sec.~D for details). The acoustic control is fast enough to keep the infidelity of the na\"ive gate below 0.05\% for GaAs QDs (${\sim}0.15\%$ for InAs QDs). As expected, the transverse noise can be neglected \cite{Merkulov2002}, as its contribution becomes considerable only for $\hbar\sigma_{\delta}>935$~neV ($T_2^{*}<1$~ns) (not shown).

Substantial literature exists on mitigating quasi-static noise effects during qubit rotations \cite{BarnesQST2022}. Composite pulses and dynamically corrected gates are compatible with restricted-angle control \cite{ZengPRA2019,KukitaPRA2022,WalelignPRA2024}. We demonstrate leading-order error cancelation with an exemplary seven-pulse sequence, found via numerical optimization (see End Matter Secs.~E and F), that maintains the $\pi/4$ inclination while only involving phase skips on the acoustic drive, with a total nutation of 3.785~$\pi$, only $1.9\times$ more than for the na\"ive gate. The red line in \subfigref{fig:fidelity}{a} shows the infidelity due to longitudinal noise for that sequence at $\omega_{\mathrm{Z}} = 44$~GHz and $\kappa=250$~ps. We find strong suppression pushing the infidelity below 0.1\% already for $T_2^{*}\approx52$~ns, leaving above order-of-magnitude headroom for current systems with cooled nuclear spins. Methods like real-time Bayesian Hamiltonian estimation \cite{Shulman2014} should enable further improvements. This demonstrates that quasi-static noise can be effectively canceled despite relatively long acoustic-gate durations.

Although we keep the trion occupation below $10^{-4}$, recombination remains the second limiting factor for fidelity. The small trion occupation incoherently returns to both spin states at a rate $1/\tau$ and forms a mixed fraction of the spin state. To determine the strength of this effect, we numerically solve Eq.~\eqref{eq:lindblad} with varying $\tau$, while neglecting the quasi-static noise and using the two-pulse sequence for simplicity.
\isubfigref{fig:fidelity}{b} shows the infidelity of spin rotation as a function of $\tau$, optimized with respect to the remaining parameters for the two Zeeman splittings. Shorter evolution accumulates less trion recombination, yielding better results for $\omega_{\mathrm{Z}} = 44$~GHz, providing faster operation. Nonetheless, both curves approach 0 with increasing $\tau$, indicating that very low infidelity is achievable with a sufficiently high $\tau$ to gate duration ratio. We mark the trion lifetimes for relevant QD systems like GaAs QDs (trion ground states with $\tau {\sim} 0.3$--1.2~ns \cite{Zhai2020,Kuster2016} and orbitally excited (envelope-dark) trion with effectively $\tau {\sim} 1$--3~ns \cite{Reindl2019,Lehner2023}), self-assembled InAs/InP QDs ($\tau {\sim} 1.5$--$2$~ns \cite{Dusanowski2018}), and vertically stacked DQD exploiting spatially indirect states with electron-hole separation \cite{Krenner2005,Bopp2023,HeynPRB2017,2505.09906}. While all the mentioned lifetimes can be extended using the Stark effect, in DQDs, the electric field allows tuning by orders of magnitude. Using $\tau = 50$~ns already provides $0.1\%$ infidelity, and much wider tuning is possible~\cite{BoyerPRL2011,HeynPRB2017}.

Lastly, we examine how the fidelity of the two-pulse gate is affected by small deviations from the optimal parameters that may occur in experiments. \isubfigref{fig:fidelity}{c} addresses laser amplitude and detuning, while in \subfigref{fig:fidelity}{d}, we vary acoustic field amplitude and phase difference. In both cases, spin rotation is robust to perturbations as fidelity remains high in wide (approx. $\pm 5\%$) ranges of parameters. Note that the color scale ends at $1\%$. The disappearance of fidelity in \subfigref{fig:fidelity}{c} for $\hbar\Delta\approx0.819~\mathrm{meV}= 4.5\hbar\omega_{\mathrm{ac}}$ is caused by an accidental resonance leading to the phonon-assisted trion excitation. We do not show the dependence on $\omega_{\mathrm{ac}}$, as it can be precisely set with no fluctuations \cite{Weiss2021}, and its deviations are mainly equivalent to $\delta_z$.

{\it Conclusions and outlook---}We have proposed a hybrid acousto-optical method of single spin control in a QD, enabling the spin-phonon interface in the system that practically lacked it. 
Combining acoustic modulation with off-resonant optical driving enables coherent spin transitions. The operation can be fast enough to allow efficient cancellation of quasi-static noise errors in QDs with cooled nuclear spins. The remaining important limitation to fidelity, caused by the finite trion lifetime, can be reduced by large Zeeman splittings, resulting in shorter evolution. This requires high acoustic frequencies, but even for the currently achievable 44~GHz we show 99.9\% fidelity for a DQD. Acoustic spin rotation has a direct advantage over microwave-based methods. Due to their low sound velocity, acoustic waves are four orders of magnitude shorter, enabling seamless on-chip integration. Introducing acoustic coupling for single spin states to QDs has broader implications. With established interfaces to light, microwaves, and nuclear spin registers, now extended to phonons, QDs could become fundamental elements in hybrid quantum architectures. This capability could open pathways for controllable spin-phonon entanglement and acoustic state transfer. Moreover, the simultaneous coupling of QD spins to multiple physical fields creates the potential for QDs to become versatile transducers, facilitating quantum information transfer not only across the electromagnetic spectrum but also between distinct physical domains.

{\it Acknowledgments---}We thank Matthias Weiß, and Hubert J. Krenner for discussions.
M.~K.\ and P.~M.\ acknowledge support from the National Science Centre (Poland) under Grants Nos.\ 2023/49/N/ST3/03931 and 2023/50/A/ST3/00511, respectively.
M.~G.\ acknowledges the financing of the MEEDGARD project funded within the QuantERA II Program that has received funding from the European Union's Horizon 2020 research and innovation program under Grant Agreement No.\ 101017733 and the National Centre for Research and Development, Poland -- project No.\ QUANTERAII/2/56/MEEDGARD/2024. This work has been supported by a Research Group Linkage Grant of the Alexander von Humboldt-Foundation funded by the German Federal Ministry of Education and Research (BMBF). 

\bibliography{hybridSpinControl}

\begin{thebibliography}{63}%
\makeatletter
\providecommand \@ifxundefined [1]{%
 \@ifx{#1\undefined}
}%
\providecommand \@ifnum [1]{%
 \ifnum #1\expandafter \@firstoftwo
 \else \expandafter \@secondoftwo
 \fi
}%
\providecommand \@ifx [1]{%
 \ifx #1\expandafter \@firstoftwo
 \else \expandafter \@secondoftwo
 \fi
}%
\providecommand \natexlab [1]{#1}%
\providecommand \enquote  [1]{``#1''}%
\providecommand \bibnamefont  [1]{#1}%
\providecommand \bibfnamefont [1]{#1}%
\providecommand \citenamefont [1]{#1}%
\providecommand \href@noop [0]{\@secondoftwo}%
\providecommand \href [0]{\begingroup \@sanitize@url \@href}%
\providecommand \@href[1]{\@@startlink{#1}\@@href}%
\providecommand \@@href[1]{\endgroup#1\@@endlink}%
\providecommand \@sanitize@url [0]{\catcode `\\12\catcode `\$12\catcode `\&12\catcode `\#12\catcode `\^12\catcode `\_12\catcode `\%12\relax}%
\providecommand \@@startlink[1]{}%
\providecommand \@@endlink[0]{}%
\providecommand \url  [0]{\begingroup\@sanitize@url \@url }%
\providecommand \@url [1]{\endgroup\@href {#1}{\urlprefix }}%
\providecommand \urlprefix  [0]{URL }%
\providecommand \Eprint [0]{\href }%
\providecommand \doibase [0]{https://doi.org/}%
\providecommand \selectlanguage [0]{\@gobble}%
\providecommand \bibinfo  [0]{\@secondoftwo}%
\providecommand \bibfield  [0]{\@secondoftwo}%
\providecommand \translation [1]{[#1]}%
\providecommand \BibitemOpen [0]{}%
\providecommand \bibitemStop [0]{}%
\providecommand \bibitemNoStop [0]{.\EOS\space}%
\providecommand \EOS [0]{\spacefactor3000\relax}%
\providecommand \BibitemShut  [1]{\csname bibitem#1\endcsname}%
\let\auto@bib@innerbib\@empty
\bibitem [{\citenamefont {Campbell}(1989)}]{Campbell1989}%
  \BibitemOpen
  \bibfield  {author} {\bibinfo {author} {\bibfnamefont {C.}~\bibnamefont {Campbell}},\ }\href@noop {} {\emph {\bibinfo {title} {Surface Acoustic Wave Devices and their Signal Processing Applications}}}\ (\bibinfo  {publisher} {Academic Press},\ \bibinfo {year} {1989})\BibitemShut {NoStop}%
\bibitem [{\citenamefont {Hashimoto}(2000)}]{Hashimoto2000}%
  \BibitemOpen
  \bibfield  {author} {\bibinfo {author} {\bibfnamefont {K.-y.}\ \bibnamefont {Hashimoto}},\ }\href@noop {} {\emph {\bibinfo {title} {Surface acoustic wave devices in telecommunications. Modelling and simulation}}}\ (\bibinfo  {publisher} {Springer Berlin, Heidelberg},\ \bibinfo {year} {2000})\BibitemShut {NoStop}%
\bibitem [{\citenamefont {Delsing}\ \emph {et~al.}(2019)\citenamefont {Delsing}, \citenamefont {Cleland}, \citenamefont {Schuetz}, \citenamefont {Knörzer}, \citenamefont {Giedke}, \citenamefont {Cirac}, \citenamefont {Srinivasan}, \citenamefont {Wu}, \citenamefont {Balram}, \citenamefont {Bäuerle}, \citenamefont {Meunier}, \citenamefont {Ford}, \citenamefont {Santos}, \citenamefont {Cerda-Méndez}, \citenamefont {Wang}, \citenamefont {Krenner}, \citenamefont {Nysten}, \citenamefont {Weiß}, \citenamefont {Nash}, \citenamefont {Thevenard}, \citenamefont {Gourdon}, \citenamefont {Rovillain}, \citenamefont {Marangolo}, \citenamefont {Duquesne}, \citenamefont {Fischerauer}, \citenamefont {Ruile}, \citenamefont {Reiner}, \citenamefont {Paschke}, \citenamefont {Denysenko}, \citenamefont {Volkmer}, \citenamefont {Wixforth}, \citenamefont {Bruus}, \citenamefont {Wiklund}, \citenamefont {Reboud}, \citenamefont {Cooper}, \citenamefont {Fu}, \citenamefont {Brugger}, \citenamefont {Rehfeldt},\ and\ \citenamefont
  {Westerhausen}}]{Delsing2019}%
  \BibitemOpen
  \bibfield  {author} {\bibinfo {author} {\bibfnamefont {P.}~\bibnamefont {Delsing}}, \bibinfo {author} {\bibfnamefont {A.~N.}\ \bibnamefont {Cleland}}, \bibinfo {author} {\bibfnamefont {M.~J.~A.}\ \bibnamefont {Schuetz}}, \bibinfo {author} {\bibfnamefont {J.}~\bibnamefont {Knörzer}}, \bibinfo {author} {\bibfnamefont {G.}~\bibnamefont {Giedke}}, \bibinfo {author} {\bibfnamefont {J.~I.}\ \bibnamefont {Cirac}}, \bibinfo {author} {\bibfnamefont {K.}~\bibnamefont {Srinivasan}}, \bibinfo {author} {\bibfnamefont {M.}~\bibnamefont {Wu}}, \bibinfo {author} {\bibfnamefont {K.~C.}\ \bibnamefont {Balram}}, \bibinfo {author} {\bibfnamefont {C.}~\bibnamefont {Bäuerle}}, \bibinfo {author} {\bibfnamefont {T.}~\bibnamefont {Meunier}}, \bibinfo {author} {\bibfnamefont {C.~J.~B.}\ \bibnamefont {Ford}}, \bibinfo {author} {\bibfnamefont {P.~V.}\ \bibnamefont {Santos}}, \bibinfo {author} {\bibfnamefont {E.}~\bibnamefont {Cerda-Méndez}}, \bibinfo {author} {\bibfnamefont {H.}~\bibnamefont {Wang}}, \bibinfo {author}
  {\bibfnamefont {H.~J.}\ \bibnamefont {Krenner}}, \bibinfo {author} {\bibfnamefont {E.~D.~S.}\ \bibnamefont {Nysten}}, \bibinfo {author} {\bibfnamefont {M.}~\bibnamefont {Weiß}}, \bibinfo {author} {\bibfnamefont {G.~R.}\ \bibnamefont {Nash}}, \bibinfo {author} {\bibfnamefont {L.}~\bibnamefont {Thevenard}}, \bibinfo {author} {\bibfnamefont {C.}~\bibnamefont {Gourdon}}, \bibinfo {author} {\bibfnamefont {P.}~\bibnamefont {Rovillain}}, \bibinfo {author} {\bibfnamefont {M.}~\bibnamefont {Marangolo}}, \bibinfo {author} {\bibfnamefont {J.-Y.}\ \bibnamefont {Duquesne}}, \bibinfo {author} {\bibfnamefont {G.}~\bibnamefont {Fischerauer}}, \bibinfo {author} {\bibfnamefont {W.}~\bibnamefont {Ruile}}, \bibinfo {author} {\bibfnamefont {A.}~\bibnamefont {Reiner}}, \bibinfo {author} {\bibfnamefont {B.}~\bibnamefont {Paschke}}, \bibinfo {author} {\bibfnamefont {D.}~\bibnamefont {Denysenko}}, \bibinfo {author} {\bibfnamefont {D.}~\bibnamefont {Volkmer}}, \bibinfo {author} {\bibfnamefont {A.}~\bibnamefont {Wixforth}}, \bibinfo
  {author} {\bibfnamefont {H.}~\bibnamefont {Bruus}}, \bibinfo {author} {\bibfnamefont {M.}~\bibnamefont {Wiklund}}, \bibinfo {author} {\bibfnamefont {J.}~\bibnamefont {Reboud}}, \bibinfo {author} {\bibfnamefont {J.~M.}\ \bibnamefont {Cooper}}, \bibinfo {author} {\bibfnamefont {Y.}~\bibnamefont {Fu}}, \bibinfo {author} {\bibfnamefont {M.~S.}\ \bibnamefont {Brugger}}, \bibinfo {author} {\bibfnamefont {F.}~\bibnamefont {Rehfeldt}},\ and\ \bibinfo {author} {\bibfnamefont {C.}~\bibnamefont {Westerhausen}},\ }\bibfield  {title} {\bibinfo {title} {The 2019 surface acoustic waves roadmap},\ }\href {https://doi.org/10.1088/1361-6463/ab1b04} {\bibfield  {journal} {\bibinfo  {journal} {J. Phys. D: Appl. Phys.}\ }\textbf {\bibinfo {volume} {52}},\ \bibinfo {pages} {353001} (\bibinfo {year} {2019})}\BibitemShut {NoStop}%
\bibitem [{\citenamefont {Choquer}\ \emph {et~al.}(2022)\citenamefont {Choquer}, \citenamefont {Weiß}, \citenamefont {Nysten}, \citenamefont {Lienhart}, \citenamefont {Machnikowski}, \citenamefont {Wigger}, \citenamefont {Krenner},\ and\ \citenamefont {Moody}}]{Choquer2022}%
  \BibitemOpen
  \bibfield  {author} {\bibinfo {author} {\bibfnamefont {M.}~\bibnamefont {Choquer}}, \bibinfo {author} {\bibfnamefont {M.}~\bibnamefont {Weiß}}, \bibinfo {author} {\bibfnamefont {E.~D.~S.}\ \bibnamefont {Nysten}}, \bibinfo {author} {\bibfnamefont {M.}~\bibnamefont {Lienhart}}, \bibinfo {author} {\bibfnamefont {P.}~\bibnamefont {Machnikowski}}, \bibinfo {author} {\bibfnamefont {D.}~\bibnamefont {Wigger}}, \bibinfo {author} {\bibfnamefont {H.~J.}\ \bibnamefont {Krenner}},\ and\ \bibinfo {author} {\bibfnamefont {G.}~\bibnamefont {Moody}},\ }\bibfield  {title} {\bibinfo {title} {Quantum control of optically active artificial atoms with surface acoustic waves},\ }\href {https://doi.org/10.1109/TQE.2022.3204928} {\bibfield  {journal} {\bibinfo  {journal} {IEEE Trans. Quantum Eng.}\ }\textbf {\bibinfo {volume} {3}},\ \bibinfo {pages} {1} (\bibinfo {year} {2022})}\BibitemShut {NoStop}%
\bibitem [{\citenamefont {Zheng}\ \emph {et~al.}(2020)\citenamefont {Zheng}, \citenamefont {Zhou}, \citenamefont {Zeng}, \citenamefont {Liu}, \citenamefont {Shen}, \citenamefont {Yao}, \citenamefont {Chen}, \citenamefont {Wu}, \citenamefont {Xiong}, \citenamefont {Chen}, \citenamefont {Shi}, \citenamefont {Liu}, \citenamefont {Fu},\ and\ \citenamefont {Duan}}]{Zheng2020}%
  \BibitemOpen
  \bibfield  {author} {\bibinfo {author} {\bibfnamefont {J.}~\bibnamefont {Zheng}}, \bibinfo {author} {\bibfnamefont {J.}~\bibnamefont {Zhou}}, \bibinfo {author} {\bibfnamefont {P.}~\bibnamefont {Zeng}}, \bibinfo {author} {\bibfnamefont {Y.}~\bibnamefont {Liu}}, \bibinfo {author} {\bibfnamefont {Y.}~\bibnamefont {Shen}}, \bibinfo {author} {\bibfnamefont {W.}~\bibnamefont {Yao}}, \bibinfo {author} {\bibfnamefont {Z.}~\bibnamefont {Chen}}, \bibinfo {author} {\bibfnamefont {J.}~\bibnamefont {Wu}}, \bibinfo {author} {\bibfnamefont {S.}~\bibnamefont {Xiong}}, \bibinfo {author} {\bibfnamefont {Y.}~\bibnamefont {Chen}}, \bibinfo {author} {\bibfnamefont {X.}~\bibnamefont {Shi}}, \bibinfo {author} {\bibfnamefont {J.}~\bibnamefont {Liu}}, \bibinfo {author} {\bibfnamefont {Y.}~\bibnamefont {Fu}},\ and\ \bibinfo {author} {\bibfnamefont {H.}~\bibnamefont {Duan}},\ }\bibfield  {title} {\bibinfo {title} {30 {GHz} surface acoustic wave transducers with extremely high mass sensitivity},\ }\href
  {https://doi.org/10.1063/1.5142673} {\bibfield  {journal} {\bibinfo  {journal} {Appl. Phys. Lett.}\ }\textbf {\bibinfo {volume} {116}},\ \bibinfo {pages} {123502} (\bibinfo {year} {2020})}\BibitemShut {NoStop}%
\bibitem [{\citenamefont {Zhou}\ \emph {et~al.}(2023)\citenamefont {Zhou}, \citenamefont {Zhang}, \citenamefont {Liu}, \citenamefont {Zhuo}, \citenamefont {Qian}, \citenamefont {Li}, \citenamefont {Fu},\ and\ \citenamefont {Duan}}]{Zhou2023}%
  \BibitemOpen
  \bibfield  {author} {\bibinfo {author} {\bibfnamefont {J.}~\bibnamefont {Zhou}}, \bibinfo {author} {\bibfnamefont {D.}~\bibnamefont {Zhang}}, \bibinfo {author} {\bibfnamefont {Y.}~\bibnamefont {Liu}}, \bibinfo {author} {\bibfnamefont {F.}~\bibnamefont {Zhuo}}, \bibinfo {author} {\bibfnamefont {L.}~\bibnamefont {Qian}}, \bibinfo {author} {\bibfnamefont {H.}~\bibnamefont {Li}}, \bibinfo {author} {\bibfnamefont {Y.-Q.}\ \bibnamefont {Fu}},\ and\ \bibinfo {author} {\bibfnamefont {H.}~\bibnamefont {Duan}},\ }\bibfield  {title} {\bibinfo {title} {Record-breaking frequency of 44 {GHz} based on the higher order mode of surface acoustic waves with {LiNbO3/SiO2/SiC} heterostructures},\ }\href {https://doi.org/https://doi.org/10.1016/j.eng.2022.05.003} {\bibfield  {journal} {\bibinfo  {journal} {Engineering}\ }\textbf {\bibinfo {volume} {20}},\ \bibinfo {pages} {112} (\bibinfo {year} {2023})}\BibitemShut {NoStop}%
\bibitem [{\citenamefont {Priya}\ \emph {et~al.}(2023)\citenamefont {Priya}, \citenamefont {Cardozo~de Oliveira},\ and\ \citenamefont {Lanzillotti-Kimura}}]{Priya2023}%
  \BibitemOpen
  \bibfield  {author} {\bibinfo {author} {\bibnamefont {Priya}}, \bibinfo {author} {\bibfnamefont {E.~R.}\ \bibnamefont {Cardozo~de Oliveira}},\ and\ \bibinfo {author} {\bibfnamefont {N.~D.}\ \bibnamefont {Lanzillotti-Kimura}},\ }\bibfield  {title} {\bibinfo {title} {Perspectives on high-frequency nanomechanics, nanoacoustics, and nanophononics},\ }\bibfield  {journal} {\bibinfo  {journal} {Appl. Phys. Lett.}\ }\textbf {\bibinfo {volume} {122}},\ \href {https://doi.org/10.1063/5.0142925} {10.1063/5.0142925} (\bibinfo {year} {2023})\BibitemShut {NoStop}%
\bibitem [{\citenamefont {Sch\"ulein}\ \emph {et~al.}(2013)\citenamefont {Sch\"ulein}, \citenamefont {M\"uller}, \citenamefont {Bichler}, \citenamefont {Koblm\"uller}, \citenamefont {Finley}, \citenamefont {Wixforth},\ and\ \citenamefont {Krenner}}]{Schulein2013}%
  \BibitemOpen
  \bibfield  {author} {\bibinfo {author} {\bibfnamefont {F.~J.~R.}\ \bibnamefont {Sch\"ulein}}, \bibinfo {author} {\bibfnamefont {K.}~\bibnamefont {M\"uller}}, \bibinfo {author} {\bibfnamefont {M.}~\bibnamefont {Bichler}}, \bibinfo {author} {\bibfnamefont {G.}~\bibnamefont {Koblm\"uller}}, \bibinfo {author} {\bibfnamefont {J.~J.}\ \bibnamefont {Finley}}, \bibinfo {author} {\bibfnamefont {A.}~\bibnamefont {Wixforth}},\ and\ \bibinfo {author} {\bibfnamefont {H.~J.}\ \bibnamefont {Krenner}},\ }\bibfield  {title} {\bibinfo {title} {Acoustically regulated carrier injection into a single optically active quantum dot},\ }\href {https://doi.org/10.1103/PhysRevB.88.085307} {\bibfield  {journal} {\bibinfo  {journal} {Phys. Rev. B}\ }\textbf {\bibinfo {volume} {88}},\ \bibinfo {pages} {085307} (\bibinfo {year} {2013})}\BibitemShut {NoStop}%
\bibitem [{\citenamefont {Wei{\ss}}\ \emph {et~al.}(2021)\citenamefont {Wei{\ss}}, \citenamefont {Wigger}, \citenamefont {N\"{a}gele}, \citenamefont {M\"{u}ller}, \citenamefont {Finley}, \citenamefont {Kuhn}, \citenamefont {Machnikowski},\ and\ \citenamefont {Krenner}}]{Weiss2021}%
  \BibitemOpen
  \bibfield  {author} {\bibinfo {author} {\bibfnamefont {M.}~\bibnamefont {Wei{\ss}}}, \bibinfo {author} {\bibfnamefont {D.}~\bibnamefont {Wigger}}, \bibinfo {author} {\bibfnamefont {M.}~\bibnamefont {N\"{a}gele}}, \bibinfo {author} {\bibfnamefont {K.}~\bibnamefont {M\"{u}ller}}, \bibinfo {author} {\bibfnamefont {J.~J.}\ \bibnamefont {Finley}}, \bibinfo {author} {\bibfnamefont {T.}~\bibnamefont {Kuhn}}, \bibinfo {author} {\bibfnamefont {P.}~\bibnamefont {Machnikowski}},\ and\ \bibinfo {author} {\bibfnamefont {H.~J.}\ \bibnamefont {Krenner}},\ }\bibfield  {title} {\bibinfo {title} {Optomechanical wave mixing by a single quantum dot},\ }\href {https://doi.org/10.1364/OPTICA.412201} {\bibfield  {journal} {\bibinfo  {journal} {Optica}\ }\textbf {\bibinfo {volume} {8}},\ \bibinfo {pages} {291} (\bibinfo {year} {2021})}\BibitemShut {NoStop}%
\bibitem [{\citenamefont {B{\"u}hler}\ \emph {et~al.}(2022)\citenamefont {B{\"u}hler}, \citenamefont {Wei{\ss}}, \citenamefont {Crespo-Poveda}, \citenamefont {Nysten}, \citenamefont {Finley}, \citenamefont {M{\"u}ller}, \citenamefont {Santos}, \citenamefont {de~Lima},\ and\ \citenamefont {Krenner}}]{Buhler2022}%
  \BibitemOpen
  \bibfield  {author} {\bibinfo {author} {\bibfnamefont {D.~D.}\ \bibnamefont {B{\"u}hler}}, \bibinfo {author} {\bibfnamefont {M.}~\bibnamefont {Wei{\ss}}}, \bibinfo {author} {\bibfnamefont {A.}~\bibnamefont {Crespo-Poveda}}, \bibinfo {author} {\bibfnamefont {E.~D.~S.}\ \bibnamefont {Nysten}}, \bibinfo {author} {\bibfnamefont {J.~J.}\ \bibnamefont {Finley}}, \bibinfo {author} {\bibfnamefont {K.}~\bibnamefont {M{\"u}ller}}, \bibinfo {author} {\bibfnamefont {P.~V.}\ \bibnamefont {Santos}}, \bibinfo {author} {\bibfnamefont {M.~M.}\ \bibnamefont {de~Lima}},\ and\ \bibinfo {author} {\bibfnamefont {H.~J.}\ \bibnamefont {Krenner}},\ }\bibfield  {title} {\bibinfo {title} {On-chip generation and dynamic piezo-optomechanical rotation of single photons},\ }\href {https://doi.org/10.1038/s41467-022-34372-9} {\bibfield  {journal} {\bibinfo  {journal} {Nat. Commun.}\ }\textbf {\bibinfo {volume} {13}},\ \bibinfo {pages} {6998} (\bibinfo {year} {2022})}\BibitemShut {NoStop}%
\bibitem [{\citenamefont {Maity}\ \emph {et~al.}(2020)\citenamefont {Maity}, \citenamefont {Shao}, \citenamefont {Bogdanovi{\'{c}}}, \citenamefont {Meesala}, \citenamefont {Sohn}, \citenamefont {Sinclair}, \citenamefont {Pingault}, \citenamefont {Chalupnik}, \citenamefont {Chia}, \citenamefont {Zheng}, \citenamefont {Lai},\ and\ \citenamefont {Lon{\v{c}}ar}}]{Maity2020}%
  \BibitemOpen
  \bibfield  {author} {\bibinfo {author} {\bibfnamefont {S.}~\bibnamefont {Maity}}, \bibinfo {author} {\bibfnamefont {L.}~\bibnamefont {Shao}}, \bibinfo {author} {\bibfnamefont {S.}~\bibnamefont {Bogdanovi{\'{c}}}}, \bibinfo {author} {\bibfnamefont {S.}~\bibnamefont {Meesala}}, \bibinfo {author} {\bibfnamefont {Y.-I.}\ \bibnamefont {Sohn}}, \bibinfo {author} {\bibfnamefont {N.}~\bibnamefont {Sinclair}}, \bibinfo {author} {\bibfnamefont {B.}~\bibnamefont {Pingault}}, \bibinfo {author} {\bibfnamefont {M.}~\bibnamefont {Chalupnik}}, \bibinfo {author} {\bibfnamefont {C.}~\bibnamefont {Chia}}, \bibinfo {author} {\bibfnamefont {L.}~\bibnamefont {Zheng}}, \bibinfo {author} {\bibfnamefont {K.}~\bibnamefont {Lai}},\ and\ \bibinfo {author} {\bibfnamefont {M.}~\bibnamefont {Lon{\v{c}}ar}},\ }\bibfield  {title} {\bibinfo {title} {Coherent acoustic control of a single silicon vacancy spin in diamond},\ }\href {https://doi.org/10.1038/s41467-019-13822-x} {\bibfield  {journal} {\bibinfo  {journal} {Nat. Commun.}\ }\textbf
  {\bibinfo {volume} {11}},\ \bibinfo {pages} {193} (\bibinfo {year} {2020})}\BibitemShut {NoStop}%
\bibitem [{\citenamefont {Lodahl}\ \emph {et~al.}(2015)\citenamefont {Lodahl}, \citenamefont {Mahmoodian},\ and\ \citenamefont {Stobbe}}]{Lodahl2015}%
  \BibitemOpen
  \bibfield  {author} {\bibinfo {author} {\bibfnamefont {P.}~\bibnamefont {Lodahl}}, \bibinfo {author} {\bibfnamefont {S.}~\bibnamefont {Mahmoodian}},\ and\ \bibinfo {author} {\bibfnamefont {S.}~\bibnamefont {Stobbe}},\ }\bibfield  {title} {\bibinfo {title} {{Interfacing single photons and single quantum dots with photonic nanostructures}},\ }\href {https://doi.org/10.1103/RevModPhys.87.347} {\bibfield  {journal} {\bibinfo  {journal} {Rev. Mod. Phys.}\ }\textbf {\bibinfo {volume} {87}},\ \bibinfo {pages} {347} (\bibinfo {year} {2015})}\BibitemShut {NoStop}%
\bibitem [{\citenamefont {Somaschi}\ \emph {et~al.}(2016)\citenamefont {Somaschi}, \citenamefont {Giesz}, \citenamefont {{De Santis}}, \citenamefont {Loredo}, \citenamefont {Almeida}, \citenamefont {Hornecker}, \citenamefont {Portalupi}, \citenamefont {Grange}, \citenamefont {Ant{\'{o}}n}, \citenamefont {Demory}, \citenamefont {G{\'{o}}mez}, \citenamefont {Sagnes}, \citenamefont {Lanzillotti-Kimura}, \citenamefont {Lema{\'{i}}tre}, \citenamefont {Auffeves}, \citenamefont {White}, \citenamefont {Lanco},\ and\ \citenamefont {Senellart}}]{Somaschi2016}%
  \BibitemOpen
  \bibfield  {author} {\bibinfo {author} {\bibfnamefont {N.}~\bibnamefont {Somaschi}}, \bibinfo {author} {\bibfnamefont {V.}~\bibnamefont {Giesz}}, \bibinfo {author} {\bibfnamefont {L.}~\bibnamefont {{De Santis}}}, \bibinfo {author} {\bibfnamefont {J.~C.}\ \bibnamefont {Loredo}}, \bibinfo {author} {\bibfnamefont {M.~P.}\ \bibnamefont {Almeida}}, \bibinfo {author} {\bibfnamefont {G.}~\bibnamefont {Hornecker}}, \bibinfo {author} {\bibfnamefont {S.~L.}\ \bibnamefont {Portalupi}}, \bibinfo {author} {\bibfnamefont {T.}~\bibnamefont {Grange}}, \bibinfo {author} {\bibfnamefont {C.}~\bibnamefont {Ant{\'{o}}n}}, \bibinfo {author} {\bibfnamefont {J.}~\bibnamefont {Demory}}, \bibinfo {author} {\bibfnamefont {C.}~\bibnamefont {G{\'{o}}mez}}, \bibinfo {author} {\bibfnamefont {I.}~\bibnamefont {Sagnes}}, \bibinfo {author} {\bibfnamefont {N.~D.}\ \bibnamefont {Lanzillotti-Kimura}}, \bibinfo {author} {\bibfnamefont {A.}~\bibnamefont {Lema{\'{i}}tre}}, \bibinfo {author} {\bibfnamefont {A.}~\bibnamefont {Auffeves}}, \bibinfo
  {author} {\bibfnamefont {A.~G.}\ \bibnamefont {White}}, \bibinfo {author} {\bibfnamefont {L.}~\bibnamefont {Lanco}},\ and\ \bibinfo {author} {\bibfnamefont {P.}~\bibnamefont {Senellart}},\ }\bibfield  {title} {\bibinfo {title} {{Near-optimal single-photon sources in the solid state}},\ }\href {https://doi.org/10.1038/nphoton.2016.23} {\bibfield  {journal} {\bibinfo  {journal} {Nat. Photonics}\ }\textbf {\bibinfo {volume} {10}},\ \bibinfo {pages} {340} (\bibinfo {year} {2016})}\BibitemShut {NoStop}%
\bibitem [{\citenamefont {Tomm}\ \emph {et~al.}(2021)\citenamefont {Tomm}, \citenamefont {Javadi}, \citenamefont {Antoniadis}, \citenamefont {Najer}, \citenamefont {L{\"{o}}bl}, \citenamefont {Korsch}, \citenamefont {Schott}, \citenamefont {Valentin}, \citenamefont {Wieck}, \citenamefont {Ludwig},\ and\ \citenamefont {Warburton}}]{Tomm2021}%
  \BibitemOpen
  \bibfield  {author} {\bibinfo {author} {\bibfnamefont {N.}~\bibnamefont {Tomm}}, \bibinfo {author} {\bibfnamefont {A.}~\bibnamefont {Javadi}}, \bibinfo {author} {\bibfnamefont {N.~O.}\ \bibnamefont {Antoniadis}}, \bibinfo {author} {\bibfnamefont {D.}~\bibnamefont {Najer}}, \bibinfo {author} {\bibfnamefont {M.~C.}\ \bibnamefont {L{\"{o}}bl}}, \bibinfo {author} {\bibfnamefont {A.~R.}\ \bibnamefont {Korsch}}, \bibinfo {author} {\bibfnamefont {R.}~\bibnamefont {Schott}}, \bibinfo {author} {\bibfnamefont {S.~R.}\ \bibnamefont {Valentin}}, \bibinfo {author} {\bibfnamefont {A.~D.}\ \bibnamefont {Wieck}}, \bibinfo {author} {\bibfnamefont {A.}~\bibnamefont {Ludwig}},\ and\ \bibinfo {author} {\bibfnamefont {R.~J.}\ \bibnamefont {Warburton}},\ }\bibfield  {title} {\bibinfo {title} {{A bright and fast source of coherent single photons}},\ }\href {https://doi.org/10.1038/s41565-020-00831-x} {\bibfield  {journal} {\bibinfo  {journal} {Nat. Nanotechnol.}\ }\textbf {\bibinfo {volume} {16}},\ \bibinfo {pages} {399}
  (\bibinfo {year} {2021})}\BibitemShut {NoStop}%
\bibitem [{\citenamefont {Y{\i}lmaz}\ \emph {et~al.}(2010)\citenamefont {Y{\i}lmaz}, \citenamefont {Fallahi},\ and\ \citenamefont {Imamo\u{g}lu}}]{Yilmaz2010}%
  \BibitemOpen
  \bibfield  {author} {\bibinfo {author} {\bibfnamefont {S.~T.}\ \bibnamefont {Y{\i}lmaz}}, \bibinfo {author} {\bibfnamefont {P.}~\bibnamefont {Fallahi}},\ and\ \bibinfo {author} {\bibfnamefont {A.}~\bibnamefont {Imamo\u{g}lu}},\ }\bibfield  {title} {\bibinfo {title} {Quantum-dot-spin single-photon interface},\ }\href {https://doi.org/10.1103/PhysRevLett.105.033601} {\bibfield  {journal} {\bibinfo  {journal} {Phys. Rev. Lett.}\ }\textbf {\bibinfo {volume} {105}},\ \bibinfo {pages} {033601} (\bibinfo {year} {2010})}\BibitemShut {NoStop}%
\bibitem [{\citenamefont {Delley}\ \emph {et~al.}(2017)\citenamefont {Delley}, \citenamefont {Kroner}, \citenamefont {Faelt}, \citenamefont {Wegscheider},\ and\ \citenamefont {\ifmmode \dot{I}\else \.{I}\fi{}mamo\ifmmode~\breve{g}\else \u{g}\fi{}lu}}]{Delley2017}%
  \BibitemOpen
  \bibfield  {author} {\bibinfo {author} {\bibfnamefont {Y.~L.}\ \bibnamefont {Delley}}, \bibinfo {author} {\bibfnamefont {M.}~\bibnamefont {Kroner}}, \bibinfo {author} {\bibfnamefont {S.}~\bibnamefont {Faelt}}, \bibinfo {author} {\bibfnamefont {W.}~\bibnamefont {Wegscheider}},\ and\ \bibinfo {author} {\bibfnamefont {A.}~\bibnamefont {\ifmmode \dot{I}\else \.{I}\fi{}mamo\ifmmode~\breve{g}\else \u{g}\fi{}lu}},\ }\bibfield  {title} {\bibinfo {title} {Deterministic entanglement between a propagating photon and a singlet-triplet qubit in an optically active quantum dot molecule},\ }\href {https://doi.org/10.1103/PhysRevB.96.241410} {\bibfield  {journal} {\bibinfo  {journal} {Phys. Rev. B}\ }\textbf {\bibinfo {volume} {96}},\ \bibinfo {pages} {241410} (\bibinfo {year} {2017})}\BibitemShut {NoStop}%
\bibitem [{\citenamefont {Lindner}\ and\ \citenamefont {Rudolph}(2009)}]{Lindner2009}%
  \BibitemOpen
  \bibfield  {author} {\bibinfo {author} {\bibfnamefont {N.~H.}\ \bibnamefont {Lindner}}\ and\ \bibinfo {author} {\bibfnamefont {T.}~\bibnamefont {Rudolph}},\ }\bibfield  {title} {\bibinfo {title} {Proposal for pulsed on-demand sources of photonic cluster state strings},\ }\href {https://doi.org/10.1103/PhysRevLett.103.113602} {\bibfield  {journal} {\bibinfo  {journal} {Phys. Rev. Lett.}\ }\textbf {\bibinfo {volume} {103}},\ \bibinfo {pages} {113602} (\bibinfo {year} {2009})}\BibitemShut {NoStop}%
\bibitem [{\citenamefont {Economou}\ \emph {et~al.}(2010)\citenamefont {Economou}, \citenamefont {Lindner},\ and\ \citenamefont {Rudolph}}]{Economou2010}%
  \BibitemOpen
  \bibfield  {author} {\bibinfo {author} {\bibfnamefont {S.~E.}\ \bibnamefont {Economou}}, \bibinfo {author} {\bibfnamefont {N.}~\bibnamefont {Lindner}},\ and\ \bibinfo {author} {\bibfnamefont {T.}~\bibnamefont {Rudolph}},\ }\bibfield  {title} {\bibinfo {title} {Optically generated 2-dimensional photonic cluster state from coupled quantum dots},\ }\href {https://doi.org/10.1103/PhysRevLett.105.093601} {\bibfield  {journal} {\bibinfo  {journal} {Phys. Rev. Lett.}\ }\textbf {\bibinfo {volume} {105}},\ \bibinfo {pages} {093601} (\bibinfo {year} {2010})}\BibitemShut {NoStop}%
\bibitem [{\citenamefont {Michaels}\ \emph {et~al.}(2021)\citenamefont {Michaels}, \citenamefont {Arjona~Mart{\'{i}}nez}, \citenamefont {Debroux}, \citenamefont {Parker}, \citenamefont {Stramma}, \citenamefont {Huber}, \citenamefont {Purser}, \citenamefont {Atat{\"{u}}re},\ and\ \citenamefont {Gangloff}}]{Michaels2021}%
  \BibitemOpen
  \bibfield  {author} {\bibinfo {author} {\bibfnamefont {C.~P.}\ \bibnamefont {Michaels}}, \bibinfo {author} {\bibfnamefont {J.}~\bibnamefont {Arjona~Mart{\'{i}}nez}}, \bibinfo {author} {\bibfnamefont {R.}~\bibnamefont {Debroux}}, \bibinfo {author} {\bibfnamefont {R.~A.}\ \bibnamefont {Parker}}, \bibinfo {author} {\bibfnamefont {A.~M.}\ \bibnamefont {Stramma}}, \bibinfo {author} {\bibfnamefont {L.~I.}\ \bibnamefont {Huber}}, \bibinfo {author} {\bibfnamefont {C.~M.}\ \bibnamefont {Purser}}, \bibinfo {author} {\bibfnamefont {M.}~\bibnamefont {Atat{\"{u}}re}},\ and\ \bibinfo {author} {\bibfnamefont {D.~A.}\ \bibnamefont {Gangloff}},\ }\bibfield  {title} {\bibinfo {title} {Multidimensional cluster states using a single spin-photon interface coupled strongly to an intrinsic nuclear register},\ }\href {https://doi.org/10.22331/q-2021-10-19-565} {\bibfield  {journal} {\bibinfo  {journal} {{Quantum}}\ }\textbf {\bibinfo {volume} {5}},\ \bibinfo {pages} {565} (\bibinfo {year} {2021})}\BibitemShut {NoStop}%
\bibitem [{\citenamefont {Cogan}\ \emph {et~al.}(2023)\citenamefont {Cogan}, \citenamefont {Su}, \citenamefont {Kenneth},\ and\ \citenamefont {Gershoni}}]{Cogan2023}%
  \BibitemOpen
  \bibfield  {author} {\bibinfo {author} {\bibfnamefont {D.}~\bibnamefont {Cogan}}, \bibinfo {author} {\bibfnamefont {Z.-E.}\ \bibnamefont {Su}}, \bibinfo {author} {\bibfnamefont {O.}~\bibnamefont {Kenneth}},\ and\ \bibinfo {author} {\bibfnamefont {D.}~\bibnamefont {Gershoni}},\ }\bibfield  {title} {\bibinfo {title} {Deterministic generation of indistinguishable photons in a cluster state},\ }\href {https://doi.org/10.1038/s41566-022-01152-2} {\bibfield  {journal} {\bibinfo  {journal} {Nat. Photonics}\ }\textbf {\bibinfo {volume} {17}},\ \bibinfo {pages} {324} (\bibinfo {year} {2023})}\BibitemShut {NoStop}%
\bibitem [{\citenamefont {Chekhovich}\ \emph {et~al.}(2020)\citenamefont {Chekhovich}, \citenamefont {da~Silva},\ and\ \citenamefont {Rastelli}}]{Chekhovich2020}%
  \BibitemOpen
  \bibfield  {author} {\bibinfo {author} {\bibfnamefont {E.~A.}\ \bibnamefont {Chekhovich}}, \bibinfo {author} {\bibfnamefont {S.~F.~C.}\ \bibnamefont {da~Silva}},\ and\ \bibinfo {author} {\bibfnamefont {A.}~\bibnamefont {Rastelli}},\ }\bibfield  {title} {\bibinfo {title} {Nuclear spin quantum register in an optically active semiconductor quantum dot},\ }\href {https://doi.org/10.1038/s41565-020-0769-3} {\bibfield  {journal} {\bibinfo  {journal} {Nat. Nanotech.}\ }\textbf {\bibinfo {volume} {15}},\ \bibinfo {pages} {999} (\bibinfo {year} {2020})}\BibitemShut {NoStop}%
\bibitem [{\citenamefont {Shofer}\ \emph {et~al.}(2025)\citenamefont {Shofer}, \citenamefont {Zaporski}, \citenamefont {Hayhurst~Appel}, \citenamefont {Manna}, \citenamefont {Covre~da Silva}, \citenamefont {Ghorbal}, \citenamefont {Haeusler}, \citenamefont {Rastelli}, \citenamefont {Le~Gall}, \citenamefont {Gawe\l{}czyk}, \citenamefont {Atat\"ure},\ and\ \citenamefont {Gangloff}}]{Shofer2025}%
  \BibitemOpen
  \bibfield  {author} {\bibinfo {author} {\bibfnamefont {N.}~\bibnamefont {Shofer}}, \bibinfo {author} {\bibfnamefont {L.}~\bibnamefont {Zaporski}}, \bibinfo {author} {\bibfnamefont {M.}~\bibnamefont {Hayhurst~Appel}}, \bibinfo {author} {\bibfnamefont {S.}~\bibnamefont {Manna}}, \bibinfo {author} {\bibfnamefont {S.}~\bibnamefont {Covre~da Silva}}, \bibinfo {author} {\bibfnamefont {A.}~\bibnamefont {Ghorbal}}, \bibinfo {author} {\bibfnamefont {U.}~\bibnamefont {Haeusler}}, \bibinfo {author} {\bibfnamefont {A.}~\bibnamefont {Rastelli}}, \bibinfo {author} {\bibfnamefont {C.}~\bibnamefont {Le~Gall}}, \bibinfo {author} {\bibfnamefont {M.}~\bibnamefont {Gawe\l{}czyk}}, \bibinfo {author} {\bibfnamefont {M.}~\bibnamefont {Atat\"ure}},\ and\ \bibinfo {author} {\bibfnamefont {D.~A.}\ \bibnamefont {Gangloff}},\ }\bibfield  {title} {\bibinfo {title} {Tuning the coherent interaction of an electron qubit and a nuclear magnon},\ }\href {https://doi.org/10.1103/PhysRevX.15.021004} {\bibfield  {journal} {\bibinfo  {journal}
  {Phys. Rev. X}\ }\textbf {\bibinfo {volume} {15}},\ \bibinfo {pages} {021004} (\bibinfo {year} {2025})}\BibitemShut {NoStop}%
\bibitem [{\citenamefont {Appel}\ \emph {et~al.}(2025)\citenamefont {Appel}, \citenamefont {Ghorbal}, \citenamefont {Shofer}, \citenamefont {Zaporski}, \citenamefont {Manna}, \citenamefont {da~Silva}, \citenamefont {Haeusler}, \citenamefont {Le~Gall}, \citenamefont {Rastelli}, \citenamefont {Gangloff},\ and\ \citenamefont {Atatüre}}]{Appel2025}%
  \BibitemOpen
  \bibfield  {author} {\bibinfo {author} {\bibfnamefont {M.~H.}\ \bibnamefont {Appel}}, \bibinfo {author} {\bibfnamefont {A.}~\bibnamefont {Ghorbal}}, \bibinfo {author} {\bibfnamefont {N.}~\bibnamefont {Shofer}}, \bibinfo {author} {\bibfnamefont {L.}~\bibnamefont {Zaporski}}, \bibinfo {author} {\bibfnamefont {S.}~\bibnamefont {Manna}}, \bibinfo {author} {\bibfnamefont {S.~F.~C.}\ \bibnamefont {da~Silva}}, \bibinfo {author} {\bibfnamefont {U.}~\bibnamefont {Haeusler}}, \bibinfo {author} {\bibfnamefont {C.}~\bibnamefont {Le~Gall}}, \bibinfo {author} {\bibfnamefont {A.}~\bibnamefont {Rastelli}}, \bibinfo {author} {\bibfnamefont {D.~A.}\ \bibnamefont {Gangloff}},\ and\ \bibinfo {author} {\bibfnamefont {M.}~\bibnamefont {Atatüre}},\ }\bibfield  {title} {\bibinfo {title} {A many-body quantum register for a spin qubit},\ }\href {https://doi.org/10.1038/s41567-024-02746-z} {\bibfield  {journal} {\bibinfo  {journal} {Nat. Phys.}\ }\textbf {\bibinfo {volume} {21}},\ \bibinfo {pages} {368} (\bibinfo {year}
  {2025})}\BibitemShut {NoStop}%
\bibitem [{\citenamefont {Holewa}\ \emph {et~al.}(2024)\citenamefont {Holewa}, \citenamefont {Vajner}, \citenamefont {Zi{\k{e}}ba-Ost{\'o}j}, \citenamefont {Wasiluk}, \citenamefont {Ga{\'a}l}, \citenamefont {Sakanas}, \citenamefont {Burakowski}, \citenamefont {Mrowi{\'{n}}ski}, \citenamefont {Krajnik}, \citenamefont {Xiong}, \citenamefont {Yvind}, \citenamefont {Gregersen}, \citenamefont {Musia{\l}}, \citenamefont {Huck}, \citenamefont {Heindel}, \citenamefont {Syperek},\ and\ \citenamefont {Semenova}}]{Holewa2024}%
  \BibitemOpen
  \bibfield  {author} {\bibinfo {author} {\bibfnamefont {P.}~\bibnamefont {Holewa}}, \bibinfo {author} {\bibfnamefont {D.~A.}\ \bibnamefont {Vajner}}, \bibinfo {author} {\bibfnamefont {E.}~\bibnamefont {Zi{\k{e}}ba-Ost{\'o}j}}, \bibinfo {author} {\bibfnamefont {M.}~\bibnamefont {Wasiluk}}, \bibinfo {author} {\bibfnamefont {B.}~\bibnamefont {Ga{\'a}l}}, \bibinfo {author} {\bibfnamefont {A.}~\bibnamefont {Sakanas}}, \bibinfo {author} {\bibfnamefont {M.}~\bibnamefont {Burakowski}}, \bibinfo {author} {\bibfnamefont {P.}~\bibnamefont {Mrowi{\'{n}}ski}}, \bibinfo {author} {\bibfnamefont {B.}~\bibnamefont {Krajnik}}, \bibinfo {author} {\bibfnamefont {M.}~\bibnamefont {Xiong}}, \bibinfo {author} {\bibfnamefont {K.}~\bibnamefont {Yvind}}, \bibinfo {author} {\bibfnamefont {N.}~\bibnamefont {Gregersen}}, \bibinfo {author} {\bibfnamefont {A.}~\bibnamefont {Musia{\l}}}, \bibinfo {author} {\bibfnamefont {A.}~\bibnamefont {Huck}}, \bibinfo {author} {\bibfnamefont {T.}~\bibnamefont {Heindel}}, \bibinfo {author}
  {\bibfnamefont {M.}~\bibnamefont {Syperek}},\ and\ \bibinfo {author} {\bibfnamefont {E.}~\bibnamefont {Semenova}},\ }\bibfield  {title} {\bibinfo {title} {High-throughput quantum photonic devices emitting indistinguishable photons in the telecom {C}-band},\ }\href {https://doi.org/10.1038/s41467-024-47551-7} {\bibfield  {journal} {\bibinfo  {journal} {Nat. Commun.}\ }\textbf {\bibinfo {volume} {15}},\ \bibinfo {pages} {3358} (\bibinfo {year} {2024})}\BibitemShut {NoStop}%
\bibitem [{\citenamefont {Holewa}\ \emph {et~al.}(2022)\citenamefont {Holewa}, \citenamefont {Sakanas}, \citenamefont {G{\"u}r}, \citenamefont {Mrowiński}, \citenamefont {Huck}, \citenamefont {Wang}, \citenamefont {Musiał}, \citenamefont {Yvind}, \citenamefont {Gregersen}, \citenamefont {Syperek},\ and\ \citenamefont {Semenova}}]{Holewa2022}%
  \BibitemOpen
  \bibfield  {author} {\bibinfo {author} {\bibfnamefont {P.}~\bibnamefont {Holewa}}, \bibinfo {author} {\bibfnamefont {A.}~\bibnamefont {Sakanas}}, \bibinfo {author} {\bibfnamefont {U.~M.}\ \bibnamefont {G{\"u}r}}, \bibinfo {author} {\bibfnamefont {P.}~\bibnamefont {Mrowiński}}, \bibinfo {author} {\bibfnamefont {A.}~\bibnamefont {Huck}}, \bibinfo {author} {\bibfnamefont {B.-Y.}\ \bibnamefont {Wang}}, \bibinfo {author} {\bibfnamefont {A.}~\bibnamefont {Musiał}}, \bibinfo {author} {\bibfnamefont {K.}~\bibnamefont {Yvind}}, \bibinfo {author} {\bibfnamefont {N.}~\bibnamefont {Gregersen}}, \bibinfo {author} {\bibfnamefont {M.}~\bibnamefont {Syperek}},\ and\ \bibinfo {author} {\bibfnamefont {E.}~\bibnamefont {Semenova}},\ }\bibfield  {title} {\bibinfo {title} {Bright quantum dot single-photon emitters at telecom bands heterogeneously integrated on {Si}},\ }\href {https://doi.org/10.1021/acsphotonics.2c00027} {\bibfield  {journal} {\bibinfo  {journal} {ACS Photonics}\ }\textbf {\bibinfo {volume} {9}},\ \bibinfo
  {pages} {2273} (\bibinfo {year} {2022})}\BibitemShut {NoStop}%
\bibitem [{\citenamefont {Deutsch}\ \emph {et~al.}(2023)\citenamefont {Deutsch}, \citenamefont {Buchholz}, \citenamefont {Zolatanosha}, \citenamefont {J\"{o}ns},\ and\ \citenamefont {Reuter}}]{Deutsch2023}%
  \BibitemOpen
  \bibfield  {author} {\bibinfo {author} {\bibfnamefont {D.}~\bibnamefont {Deutsch}}, \bibinfo {author} {\bibfnamefont {C.}~\bibnamefont {Buchholz}}, \bibinfo {author} {\bibfnamefont {V.}~\bibnamefont {Zolatanosha}}, \bibinfo {author} {\bibfnamefont {K.~D.}\ \bibnamefont {J\"{o}ns}},\ and\ \bibinfo {author} {\bibfnamefont {D.}~\bibnamefont {Reuter}},\ }\bibfield  {title} {\bibinfo {title} {Telecom {C}-band photon emission from {(In, Ga)As} quantum dots generated by filling nanoholes in {In$_{0.52}$Al$_{0.48}$As} layers},\ }\href {https://doi.org/10.1063/5.0147281} {\bibfield  {journal} {\bibinfo  {journal} {AIP Adv.}\ }\textbf {\bibinfo {volume} {13}},\ \bibinfo {pages} {055009} (\bibinfo {year} {2023})}\BibitemShut {NoStop}%
\bibitem [{\citenamefont {Michl}\ \emph {et~al.}(2023)\citenamefont {Michl}, \citenamefont {Peniakov}, \citenamefont {Pfenning}, \citenamefont {Hilska}, \citenamefont {Chellu}, \citenamefont {Bader}, \citenamefont {Guina}, \citenamefont {H\"{o}fling}, \citenamefont {Hakkarainen},\ and\ \citenamefont {Huber‐Loyola}}]{Michl2023}%
  \BibitemOpen
  \bibfield  {author} {\bibinfo {author} {\bibfnamefont {J.}~\bibnamefont {Michl}}, \bibinfo {author} {\bibfnamefont {G.}~\bibnamefont {Peniakov}}, \bibinfo {author} {\bibfnamefont {A.}~\bibnamefont {Pfenning}}, \bibinfo {author} {\bibfnamefont {J.}~\bibnamefont {Hilska}}, \bibinfo {author} {\bibfnamefont {A.}~\bibnamefont {Chellu}}, \bibinfo {author} {\bibfnamefont {A.}~\bibnamefont {Bader}}, \bibinfo {author} {\bibfnamefont {M.}~\bibnamefont {Guina}}, \bibinfo {author} {\bibfnamefont {S.}~\bibnamefont {H\"{o}fling}}, \bibinfo {author} {\bibfnamefont {T.}~\bibnamefont {Hakkarainen}},\ and\ \bibinfo {author} {\bibfnamefont {T.}~\bibnamefont {Huber‐Loyola}},\ }\bibfield  {title} {\bibinfo {title} {Strain‐free {GaSb} quantum dots as single‐photon sources in the telecom {S}‐band},\ }\href {https://doi.org/10.1002/qute.202300180} {\bibfield  {journal} {\bibinfo  {journal} {Adv. Quantum Technol.}\ }\textbf {\bibinfo {volume} {6}},\ \bibinfo {pages} {2300180} (\bibinfo {year} {2023})}\BibitemShut {NoStop}%
\bibitem [{\citenamefont {Li}\ \emph {et~al.}(2022)\citenamefont {Li}, \citenamefont {Chen}, \citenamefont {Ng}, \citenamefont {Lim}, \citenamefont {Xue}, \citenamefont {Ho}, \citenamefont {Fu},\ and\ \citenamefont {Lee}}]{Li2022}%
  \BibitemOpen
  \bibfield  {author} {\bibinfo {author} {\bibfnamefont {N.}~\bibnamefont {Li}}, \bibinfo {author} {\bibfnamefont {G.}~\bibnamefont {Chen}}, \bibinfo {author} {\bibfnamefont {D.~K.~T.}\ \bibnamefont {Ng}}, \bibinfo {author} {\bibfnamefont {L.~W.}\ \bibnamefont {Lim}}, \bibinfo {author} {\bibfnamefont {J.}~\bibnamefont {Xue}}, \bibinfo {author} {\bibfnamefont {C.~P.}\ \bibnamefont {Ho}}, \bibinfo {author} {\bibfnamefont {Y.~H.}\ \bibnamefont {Fu}},\ and\ \bibinfo {author} {\bibfnamefont {L.~Y.~T.}\ \bibnamefont {Lee}},\ }\bibfield  {title} {\bibinfo {title} {Integrated lasers on silicon at communication wavelength: A progress review},\ }\href {https://doi.org/https://doi.org/10.1002/adom.202201008} {\bibfield  {journal} {\bibinfo  {journal} {Adv. Opt. Mater.}\ }\textbf {\bibinfo {volume} {10}},\ \bibinfo {pages} {2201008} (\bibinfo {year} {2022})}\BibitemShut {NoStop}%
\bibitem [{\citenamefont {Van~Gasse}\ \emph {et~al.}(2019)\citenamefont {Van~Gasse}, \citenamefont {Uvin}, \citenamefont {Moskalenko}, \citenamefont {Latkowski}, \citenamefont {Roelkens}, \citenamefont {Bente},\ and\ \citenamefont {Kuyken}}]{VanGasse2019}%
  \BibitemOpen
  \bibfield  {author} {\bibinfo {author} {\bibfnamefont {K.}~\bibnamefont {Van~Gasse}}, \bibinfo {author} {\bibfnamefont {S.}~\bibnamefont {Uvin}}, \bibinfo {author} {\bibfnamefont {V.}~\bibnamefont {Moskalenko}}, \bibinfo {author} {\bibfnamefont {S.}~\bibnamefont {Latkowski}}, \bibinfo {author} {\bibfnamefont {G.}~\bibnamefont {Roelkens}}, \bibinfo {author} {\bibfnamefont {E.}~\bibnamefont {Bente}},\ and\ \bibinfo {author} {\bibfnamefont {B.}~\bibnamefont {Kuyken}},\ }\bibfield  {title} {\bibinfo {title} {Recent advances in the photonic integration of mode-locked laser diodes},\ }\href {https://doi.org/10.1109/LPT.2019.2945973} {\bibfield  {journal} {\bibinfo  {journal} {IEEE Photonics Technol. Lett.}\ }\textbf {\bibinfo {volume} {31}},\ \bibinfo {pages} {1870} (\bibinfo {year} {2019})}\BibitemShut {NoStop}%
\bibitem [{\citenamefont {Schuetz}\ \emph {et~al.}(2015)\citenamefont {Schuetz}, \citenamefont {Kessler}, \citenamefont {Giedke}, \citenamefont {Vandersypen}, \citenamefont {Lukin},\ and\ \citenamefont {Cirac}}]{Schuetz2015}%
  \BibitemOpen
  \bibfield  {author} {\bibinfo {author} {\bibfnamefont {M.~J.~A.}\ \bibnamefont {Schuetz}}, \bibinfo {author} {\bibfnamefont {E.~M.}\ \bibnamefont {Kessler}}, \bibinfo {author} {\bibfnamefont {G.}~\bibnamefont {Giedke}}, \bibinfo {author} {\bibfnamefont {L.~M.~K.}\ \bibnamefont {Vandersypen}}, \bibinfo {author} {\bibfnamefont {M.~D.}\ \bibnamefont {Lukin}},\ and\ \bibinfo {author} {\bibfnamefont {J.~I.}\ \bibnamefont {Cirac}},\ }\bibfield  {title} {\bibinfo {title} {Universal quantum transducers based on surface acoustic waves},\ }\href {https://doi.org/10.1103/PhysRevX.5.031031} {\bibfield  {journal} {\bibinfo  {journal} {Phys. Rev. X}\ }\textbf {\bibinfo {volume} {5}},\ \bibinfo {pages} {031031} (\bibinfo {year} {2015})}\BibitemShut {NoStop}%
\bibitem [{\citenamefont {Sun}\ \emph {et~al.}(2012)\citenamefont {Sun}, \citenamefont {Chow}, \citenamefont {Steel}, \citenamefont {Bracker}, \citenamefont {Gammon},\ and\ \citenamefont {Sham}}]{SunPRL2012}%
  \BibitemOpen
  \bibfield  {author} {\bibinfo {author} {\bibfnamefont {B.}~\bibnamefont {Sun}}, \bibinfo {author} {\bibfnamefont {C.~M.~E.}\ \bibnamefont {Chow}}, \bibinfo {author} {\bibfnamefont {D.~G.}\ \bibnamefont {Steel}}, \bibinfo {author} {\bibfnamefont {A.~S.}\ \bibnamefont {Bracker}}, \bibinfo {author} {\bibfnamefont {D.}~\bibnamefont {Gammon}},\ and\ \bibinfo {author} {\bibfnamefont {L.~J.}\ \bibnamefont {Sham}},\ }\bibfield  {title} {\bibinfo {title} {Persistent narrowing of nuclear-spin fluctuations in {InAs} quantum dots using laser excitation},\ }\href {https://doi.org/10.1103/PhysRevLett.108.187401} {\bibfield  {journal} {\bibinfo  {journal} {Phys. Rev. Lett.}\ }\textbf {\bibinfo {volume} {108}},\ \bibinfo {pages} {187401} (\bibinfo {year} {2012})}\BibitemShut {NoStop}%
\bibitem [{\citenamefont {\'Ethier-Majcher}\ \emph {et~al.}(2017)\citenamefont {\'Ethier-Majcher}, \citenamefont {Gangloff}, \citenamefont {Stockill}, \citenamefont {Clarke}, \citenamefont {Hugues}, \citenamefont {Le~Gall},\ and\ \citenamefont {Atat\"ure}}]{EthierMajcherPRL2017}%
  \BibitemOpen
  \bibfield  {author} {\bibinfo {author} {\bibfnamefont {G.}~\bibnamefont {\'Ethier-Majcher}}, \bibinfo {author} {\bibfnamefont {D.}~\bibnamefont {Gangloff}}, \bibinfo {author} {\bibfnamefont {R.}~\bibnamefont {Stockill}}, \bibinfo {author} {\bibfnamefont {E.}~\bibnamefont {Clarke}}, \bibinfo {author} {\bibfnamefont {M.}~\bibnamefont {Hugues}}, \bibinfo {author} {\bibfnamefont {C.}~\bibnamefont {Le~Gall}},\ and\ \bibinfo {author} {\bibfnamefont {M.}~\bibnamefont {Atat\"ure}},\ }\bibfield  {title} {\bibinfo {title} {Improving a solid-state qubit through an engineered mesoscopic environment},\ }\href {https://doi.org/10.1103/PhysRevLett.119.130503} {\bibfield  {journal} {\bibinfo  {journal} {Phys. Rev. Lett.}\ }\textbf {\bibinfo {volume} {119}},\ \bibinfo {pages} {130503} (\bibinfo {year} {2017})}\BibitemShut {NoStop}%
\bibitem [{\citenamefont {Jackson}\ \emph {et~al.}(2022)\citenamefont {Jackson}, \citenamefont {Haeusler}, \citenamefont {Zaporski}, \citenamefont {Bodey}, \citenamefont {Shofer}, \citenamefont {Clarke}, \citenamefont {Hugues}, \citenamefont {Atat\"ure}, \citenamefont {Le~Gall},\ and\ \citenamefont {Gangloff}}]{JacksonPRX2022}%
  \BibitemOpen
  \bibfield  {author} {\bibinfo {author} {\bibfnamefont {D.~M.}\ \bibnamefont {Jackson}}, \bibinfo {author} {\bibfnamefont {U.}~\bibnamefont {Haeusler}}, \bibinfo {author} {\bibfnamefont {L.}~\bibnamefont {Zaporski}}, \bibinfo {author} {\bibfnamefont {J.~H.}\ \bibnamefont {Bodey}}, \bibinfo {author} {\bibfnamefont {N.}~\bibnamefont {Shofer}}, \bibinfo {author} {\bibfnamefont {E.}~\bibnamefont {Clarke}}, \bibinfo {author} {\bibfnamefont {M.}~\bibnamefont {Hugues}}, \bibinfo {author} {\bibfnamefont {M.}~\bibnamefont {Atat\"ure}}, \bibinfo {author} {\bibfnamefont {C.}~\bibnamefont {Le~Gall}},\ and\ \bibinfo {author} {\bibfnamefont {D.~A.}\ \bibnamefont {Gangloff}},\ }\bibfield  {title} {\bibinfo {title} {Optimal purification of a spin ensemble by quantum-algorithmic feedback},\ }\href {https://doi.org/10.1103/PhysRevX.12.031014} {\bibfield  {journal} {\bibinfo  {journal} {Phys. Rev. X}\ }\textbf {\bibinfo {volume} {12}},\ \bibinfo {pages} {031014} (\bibinfo {year} {2022})}\BibitemShut {NoStop}%
\bibitem [{\citenamefont {Nguyen}\ \emph {et~al.}(2023)\citenamefont {Nguyen}, \citenamefont {Spinnler}, \citenamefont {Hogg}, \citenamefont {Zhai}, \citenamefont {Javadi}, \citenamefont {Schrader}, \citenamefont {Erbe}, \citenamefont {Wyss}, \citenamefont {Ritzmann}, \citenamefont {Babin}, \citenamefont {Wieck}, \citenamefont {Ludwig},\ and\ \citenamefont {Warburton}}]{NguyenPRL2023}%
  \BibitemOpen
  \bibfield  {author} {\bibinfo {author} {\bibfnamefont {G.~N.}\ \bibnamefont {Nguyen}}, \bibinfo {author} {\bibfnamefont {C.}~\bibnamefont {Spinnler}}, \bibinfo {author} {\bibfnamefont {M.~R.}\ \bibnamefont {Hogg}}, \bibinfo {author} {\bibfnamefont {L.}~\bibnamefont {Zhai}}, \bibinfo {author} {\bibfnamefont {A.}~\bibnamefont {Javadi}}, \bibinfo {author} {\bibfnamefont {C.~A.}\ \bibnamefont {Schrader}}, \bibinfo {author} {\bibfnamefont {M.}~\bibnamefont {Erbe}}, \bibinfo {author} {\bibfnamefont {M.}~\bibnamefont {Wyss}}, \bibinfo {author} {\bibfnamefont {J.}~\bibnamefont {Ritzmann}}, \bibinfo {author} {\bibfnamefont {H.-G.}\ \bibnamefont {Babin}}, \bibinfo {author} {\bibfnamefont {A.~D.}\ \bibnamefont {Wieck}}, \bibinfo {author} {\bibfnamefont {A.}~\bibnamefont {Ludwig}},\ and\ \bibinfo {author} {\bibfnamefont {R.~J.}\ \bibnamefont {Warburton}},\ }\bibfield  {title} {\bibinfo {title} {Enhanced electron-spin coherence in a {GaAs} quantum emitter},\ }\href {https://doi.org/10.1103/PhysRevLett.131.210805}
  {\bibfield  {journal} {\bibinfo  {journal} {Phys. Rev. Lett.}\ }\textbf {\bibinfo {volume} {131}},\ \bibinfo {pages} {210805} (\bibinfo {year} {2023})}\BibitemShut {NoStop}%
\bibitem [{\citenamefont {Scully}\ and\ \citenamefont {Zubairy}(1997)}]{Scully1997}%
  \BibitemOpen
  \bibfield  {author} {\bibinfo {author} {\bibfnamefont {M.~O.}\ \bibnamefont {Scully}}\ and\ \bibinfo {author} {\bibfnamefont {M.~S.}\ \bibnamefont {Zubairy}},\ }\href@noop {} {\emph {\bibinfo {title} {Quantum Optics}}}\ (\bibinfo  {publisher} {Cambridge University Press},\ \bibinfo {year} {1997})\BibitemShut {NoStop}%
\bibitem [{\citenamefont {Golter}\ \emph {et~al.}(2016)\citenamefont {Golter}, \citenamefont {Oo}, \citenamefont {Amezcua}, \citenamefont {Lekavicius}, \citenamefont {Stewart},\ and\ \citenamefont {Wang}}]{GolterPRX2016}%
  \BibitemOpen
  \bibfield  {author} {\bibinfo {author} {\bibfnamefont {D.~A.}\ \bibnamefont {Golter}}, \bibinfo {author} {\bibfnamefont {T.}~\bibnamefont {Oo}}, \bibinfo {author} {\bibfnamefont {M.}~\bibnamefont {Amezcua}}, \bibinfo {author} {\bibfnamefont {I.}~\bibnamefont {Lekavicius}}, \bibinfo {author} {\bibfnamefont {K.~A.}\ \bibnamefont {Stewart}},\ and\ \bibinfo {author} {\bibfnamefont {H.}~\bibnamefont {Wang}},\ }\bibfield  {title} {\bibinfo {title} {Coupling a surface acoustic wave to an electron spin in diamond via a dark state},\ }\href {https://doi.org/10.1103/PhysRevX.6.041060} {\bibfield  {journal} {\bibinfo  {journal} {Phys. Rev. X}\ }\textbf {\bibinfo {volume} {6}},\ \bibinfo {pages} {041060} (\bibinfo {year} {2016})}\BibitemShut {NoStop}%
\bibitem [{\citenamefont {Nysten}(2021)}]{Nysten2021PhD}%
  \BibitemOpen
  \bibfield  {author} {\bibinfo {author} {\bibfnamefont {E.~D.~S.}\ \bibnamefont {Nysten}},\ }\emph {\bibinfo {title} {Hybrid LiNbO$_3$--{(Al)GaAs} devices for quantum dot optomechanics}},\ \href@noop {} {Ph.D. thesis},\ \bibinfo  {school} {Faculty of Mathematics, Natural Sciences, and Materials Engineering, University of Augsburg}, \bibinfo {address} {Augsburg, Germany} (\bibinfo {year} {2021})\BibitemShut {NoStop}%
\bibitem [{\citenamefont {Merkulov}\ \emph {et~al.}(2002)\citenamefont {Merkulov}, \citenamefont {Efros},\ and\ \citenamefont {Rosen}}]{Merkulov2002}%
  \BibitemOpen
  \bibfield  {author} {\bibinfo {author} {\bibfnamefont {I.~A.}\ \bibnamefont {Merkulov}}, \bibinfo {author} {\bibfnamefont {A.~L.}\ \bibnamefont {Efros}},\ and\ \bibinfo {author} {\bibfnamefont {M.}~\bibnamefont {Rosen}},\ }\bibfield  {title} {\bibinfo {title} {Electron spin relaxation by nuclei in semiconductor quantum dots},\ }\href {https://doi.org/10.1103/PhysRevB.65.205309} {\bibfield  {journal} {\bibinfo  {journal} {Phys. Rev. B}\ }\textbf {\bibinfo {volume} {65}},\ \bibinfo {pages} {205309} (\bibinfo {year} {2002})}\BibitemShut {NoStop}%
\bibitem [{\citenamefont {Nysten}\ \emph {et~al.}(2017)\citenamefont {Nysten}, \citenamefont {Huo}, \citenamefont {Yu}, \citenamefont {Song}, \citenamefont {Rastelli},\ and\ \citenamefont {Krenner}}]{Nysten2017}%
  \BibitemOpen
  \bibfield  {author} {\bibinfo {author} {\bibfnamefont {E.~D.~S.}\ \bibnamefont {Nysten}}, \bibinfo {author} {\bibfnamefont {Y.~H.}\ \bibnamefont {Huo}}, \bibinfo {author} {\bibfnamefont {H.}~\bibnamefont {Yu}}, \bibinfo {author} {\bibfnamefont {G.~F.}\ \bibnamefont {Song}}, \bibinfo {author} {\bibfnamefont {A.}~\bibnamefont {Rastelli}},\ and\ \bibinfo {author} {\bibfnamefont {H.~J.}\ \bibnamefont {Krenner}},\ }\bibfield  {title} {\bibinfo {title} {Multi-harmonic quantum dot optomechanics in fused {LiNbO$_3$}–{(Al)GaAs} hybrids},\ }\href {https://doi.org/10.1088/1361-6463/aa861a} {\bibfield  {journal} {\bibinfo  {journal} {J. Phys. D: Appl. Phys.}\ }\textbf {\bibinfo {volume} {50}},\ \bibinfo {pages} {43LT01} (\bibinfo {year} {2017})}\BibitemShut {NoStop}%
\bibitem [{\citenamefont {Kukushkin}\ \emph {et~al.}(2006)\citenamefont {Kukushkin}, \citenamefont {Smet}, \citenamefont {Lyne~Abergel}, \citenamefont {Fal'ko}, \citenamefont {Wegscheider},\ and\ \citenamefont {von Klitzing}}]{Kukushkin2006}%
  \BibitemOpen
  \bibfield  {author} {\bibinfo {author} {\bibfnamefont {I.~V.}\ \bibnamefont {Kukushkin}}, \bibinfo {author} {\bibfnamefont {J.~H.}\ \bibnamefont {Smet}}, \bibinfo {author} {\bibfnamefont {D.~S.}\ \bibnamefont {Lyne~Abergel}}, \bibinfo {author} {\bibfnamefont {V.~I.}\ \bibnamefont {Fal'ko}}, \bibinfo {author} {\bibfnamefont {W.}~\bibnamefont {Wegscheider}},\ and\ \bibinfo {author} {\bibfnamefont {K.}~\bibnamefont {von Klitzing}},\ }\bibfield  {title} {\bibinfo {title} {Detection of the electron spin resonance of two-dimensional electrons at large wave vectors},\ }\href {https://doi.org/10.1103/PhysRevLett.96.126807} {\bibfield  {journal} {\bibinfo  {journal} {Phys. Rev. Lett.}\ }\textbf {\bibinfo {volume} {96}},\ \bibinfo {pages} {126807} (\bibinfo {year} {2006})}\BibitemShut {NoStop}%
\bibitem [{\citenamefont {Ken}\ \emph {et~al.}(2025)\citenamefont {Ken}, \citenamefont {Horiachyi}, \citenamefont {Akimov}, \citenamefont {Korenev}, \citenamefont {Gusev}, \citenamefont {Litvin}, \citenamefont {Kahl}, \citenamefont {Ludwig}, \citenamefont {Spitzer}, \citenamefont {Wieck},\ and\ \citenamefont {Bayer}}]{Ken2025}%
  \BibitemOpen
  \bibfield  {author} {\bibinfo {author} {\bibfnamefont {O.}~\bibnamefont {Ken}}, \bibinfo {author} {\bibfnamefont {D.}~\bibnamefont {Horiachyi}}, \bibinfo {author} {\bibfnamefont {I.}~\bibnamefont {Akimov}}, \bibinfo {author} {\bibfnamefont {V.}~\bibnamefont {Korenev}}, \bibinfo {author} {\bibfnamefont {V.}~\bibnamefont {Gusev}}, \bibinfo {author} {\bibfnamefont {L.}~\bibnamefont {Litvin}}, \bibinfo {author} {\bibfnamefont {M.}~\bibnamefont {Kahl}}, \bibinfo {author} {\bibfnamefont {A.}~\bibnamefont {Ludwig}}, \bibinfo {author} {\bibfnamefont {N.}~\bibnamefont {Spitzer}}, \bibinfo {author} {\bibfnamefont {A.~D.}\ \bibnamefont {Wieck}},\ and\ \bibinfo {author} {\bibfnamefont {M.}~\bibnamefont {Bayer}},\ }\href@noop {} {\bibinfo {title} {High-frequency surface acoustic waves: Generation with sub-optical wavelength metal gratings and detection at the exciton resonance}} (\bibinfo {year} {2025}),\ \Eprint {https://arxiv.org/abs/arXiv:2507.11425} {arXiv:2507.11425} \BibitemShut {NoStop}%
\bibitem [{\citenamefont {Yaremkevich}\ \emph {et~al.}(2021)\citenamefont {Yaremkevich}, \citenamefont {Scherbakov}, \citenamefont {Kukhtaruk}, \citenamefont {Linnik}, \citenamefont {Khokhlov}, \citenamefont {Godejohann}, \citenamefont {Dyatlova}, \citenamefont {Nadzeyka}, \citenamefont {Pattnaik}, \citenamefont {Wang}, \citenamefont {Roy}, \citenamefont {Campion}, \citenamefont {Rushforth}, \citenamefont {Gusev}, \citenamefont {Akimov},\ and\ \citenamefont {Bayer}}]{Yaremkevich2021}%
  \BibitemOpen
  \bibfield  {author} {\bibinfo {author} {\bibfnamefont {D.~D.}\ \bibnamefont {Yaremkevich}}, \bibinfo {author} {\bibfnamefont {A.~V.}\ \bibnamefont {Scherbakov}}, \bibinfo {author} {\bibfnamefont {S.~M.}\ \bibnamefont {Kukhtaruk}}, \bibinfo {author} {\bibfnamefont {T.~L.}\ \bibnamefont {Linnik}}, \bibinfo {author} {\bibfnamefont {N.~E.}\ \bibnamefont {Khokhlov}}, \bibinfo {author} {\bibfnamefont {F.}~\bibnamefont {Godejohann}}, \bibinfo {author} {\bibfnamefont {O.~A.}\ \bibnamefont {Dyatlova}}, \bibinfo {author} {\bibfnamefont {A.}~\bibnamefont {Nadzeyka}}, \bibinfo {author} {\bibfnamefont {D.~P.}\ \bibnamefont {Pattnaik}}, \bibinfo {author} {\bibfnamefont {M.}~\bibnamefont {Wang}}, \bibinfo {author} {\bibfnamefont {S.}~\bibnamefont {Roy}}, \bibinfo {author} {\bibfnamefont {R.~P.}\ \bibnamefont {Campion}}, \bibinfo {author} {\bibfnamefont {A.~W.}\ \bibnamefont {Rushforth}}, \bibinfo {author} {\bibfnamefont {V.~E.}\ \bibnamefont {Gusev}}, \bibinfo {author} {\bibfnamefont {A.~V.}\ \bibnamefont {Akimov}},\ and\
  \bibinfo {author} {\bibfnamefont {M.}~\bibnamefont {Bayer}},\ }\bibfield  {title} {\bibinfo {title} {Protected long-distance guiding of hypersound underneath a nanocorrugated surface},\ }\href {https://doi.org/10.1021/acsnano.0c09475} {\bibfield  {journal} {\bibinfo  {journal} {ACS Nano}\ }\textbf {\bibinfo {volume} {15}},\ \bibinfo {pages} {4802} (\bibinfo {year} {2021})}\BibitemShut {NoStop}%
\bibitem [{\citenamefont {Pustiowski}\ \emph {et~al.}(2015)\citenamefont {Pustiowski}, \citenamefont {M\"{u}ller}, \citenamefont {Bichler}, \citenamefont {Koblm\"{u}ller}, \citenamefont {Finley}, \citenamefont {Wixforth},\ and\ \citenamefont {Krenner}}]{Pustiowski2015}%
  \BibitemOpen
  \bibfield  {author} {\bibinfo {author} {\bibfnamefont {J.}~\bibnamefont {Pustiowski}}, \bibinfo {author} {\bibfnamefont {K.}~\bibnamefont {M\"{u}ller}}, \bibinfo {author} {\bibfnamefont {M.}~\bibnamefont {Bichler}}, \bibinfo {author} {\bibfnamefont {G.}~\bibnamefont {Koblm\"{u}ller}}, \bibinfo {author} {\bibfnamefont {J.~J.}\ \bibnamefont {Finley}}, \bibinfo {author} {\bibfnamefont {A.}~\bibnamefont {Wixforth}},\ and\ \bibinfo {author} {\bibfnamefont {H.~J.}\ \bibnamefont {Krenner}},\ }\bibfield  {title} {\bibinfo {title} {Independent dynamic acousto-mechanical and electrostatic control of individual quantum dots in a {LiNbO3}-{GaAs} hybrid},\ }\bibfield  {journal} {\bibinfo  {journal} {Appl. Phys. Lett.}\ }\textbf {\bibinfo {volume} {106}},\ \href {https://doi.org/10.1063/1.4905477} {10.1063/1.4905477} (\bibinfo {year} {2015})\BibitemShut {NoStop}%
\bibitem [{\citenamefont {Barnes}\ \emph {et~al.}(2022)\citenamefont {Barnes}, \citenamefont {Calderon-Vargas}, \citenamefont {Dong}, \citenamefont {Li}, \citenamefont {Zeng},\ and\ \citenamefont {Zhuang}}]{BarnesQST2022}%
  \BibitemOpen
  \bibfield  {author} {\bibinfo {author} {\bibfnamefont {E.}~\bibnamefont {Barnes}}, \bibinfo {author} {\bibfnamefont {F.~A.}\ \bibnamefont {Calderon-Vargas}}, \bibinfo {author} {\bibfnamefont {W.}~\bibnamefont {Dong}}, \bibinfo {author} {\bibfnamefont {B.}~\bibnamefont {Li}}, \bibinfo {author} {\bibfnamefont {J.}~\bibnamefont {Zeng}},\ and\ \bibinfo {author} {\bibfnamefont {F.}~\bibnamefont {Zhuang}},\ }\bibfield  {title} {\bibinfo {title} {Dynamically corrected gates from geometric space curves},\ }\href {https://doi.org/10.1088/2058-9565/ac4421} {\bibfield  {journal} {\bibinfo  {journal} {Quantum Sci. Technol.}\ }\textbf {\bibinfo {volume} {7}},\ \bibinfo {pages} {023001} (\bibinfo {year} {2022})}\BibitemShut {NoStop}%
\bibitem [{\citenamefont {Zeng}\ \emph {et~al.}(2019)\citenamefont {Zeng}, \citenamefont {Yang}, \citenamefont {Dzurak},\ and\ \citenamefont {Barnes}}]{ZengPRA2019}%
  \BibitemOpen
  \bibfield  {author} {\bibinfo {author} {\bibfnamefont {J.}~\bibnamefont {Zeng}}, \bibinfo {author} {\bibfnamefont {C.~H.}\ \bibnamefont {Yang}}, \bibinfo {author} {\bibfnamefont {A.~S.}\ \bibnamefont {Dzurak}},\ and\ \bibinfo {author} {\bibfnamefont {E.}~\bibnamefont {Barnes}},\ }\bibfield  {title} {\bibinfo {title} {Geometric formalism for constructing arbitrary single-qubit dynamically corrected gates},\ }\href {https://doi.org/10.1103/PhysRevA.99.052321} {\bibfield  {journal} {\bibinfo  {journal} {Phys. Rev. A}\ }\textbf {\bibinfo {volume} {99}},\ \bibinfo {pages} {052321} (\bibinfo {year} {2019})}\BibitemShut {NoStop}%
\bibitem [{\citenamefont {Kukita}\ \emph {et~al.}(2022)\citenamefont {Kukita}, \citenamefont {Kiya},\ and\ \citenamefont {Kondo}}]{KukitaPRA2022}%
  \BibitemOpen
  \bibfield  {author} {\bibinfo {author} {\bibfnamefont {S.}~\bibnamefont {Kukita}}, \bibinfo {author} {\bibfnamefont {H.}~\bibnamefont {Kiya}},\ and\ \bibinfo {author} {\bibfnamefont {Y.}~\bibnamefont {Kondo}},\ }\bibfield  {title} {\bibinfo {title} {General off-resonance-error-robust symmetric composite pulses with three elementary operations},\ }\href {https://doi.org/10.1103/PhysRevA.106.042613} {\bibfield  {journal} {\bibinfo  {journal} {Phys. Rev. A}\ }\textbf {\bibinfo {volume} {106}},\ \bibinfo {pages} {042613} (\bibinfo {year} {2022})}\BibitemShut {NoStop}%
\bibitem [{\citenamefont {Walelign}\ \emph {et~al.}(2024)\citenamefont {Walelign}, \citenamefont {Cai}, \citenamefont {Li}, \citenamefont {Barnes},\ and\ \citenamefont {Nichol}}]{WalelignPRA2024}%
  \BibitemOpen
  \bibfield  {author} {\bibinfo {author} {\bibfnamefont {H.~Y.}\ \bibnamefont {Walelign}}, \bibinfo {author} {\bibfnamefont {X.}~\bibnamefont {Cai}}, \bibinfo {author} {\bibfnamefont {B.}~\bibnamefont {Li}}, \bibinfo {author} {\bibfnamefont {E.}~\bibnamefont {Barnes}},\ and\ \bibinfo {author} {\bibfnamefont {J.~M.}\ \bibnamefont {Nichol}},\ }\bibfield  {title} {\bibinfo {title} {Dynamically corrected gates in silicon singlet-triplet spin qubits},\ }\href {https://doi.org/10.1103/PhysRevApplied.22.064029} {\bibfield  {journal} {\bibinfo  {journal} {Phys. Rev. Appl.}\ }\textbf {\bibinfo {volume} {22}},\ \bibinfo {pages} {064029} (\bibinfo {year} {2024})}\BibitemShut {NoStop}%
\bibitem [{\citenamefont {Shulman}\ \emph {et~al.}(2014)\citenamefont {Shulman}, \citenamefont {Harvey}, \citenamefont {Nichol}, \citenamefont {Bartlett}, \citenamefont {Doherty}, \citenamefont {Umansky},\ and\ \citenamefont {Yacoby}}]{Shulman2014}%
  \BibitemOpen
  \bibfield  {author} {\bibinfo {author} {\bibfnamefont {M.~D.}\ \bibnamefont {Shulman}}, \bibinfo {author} {\bibfnamefont {S.~P.}\ \bibnamefont {Harvey}}, \bibinfo {author} {\bibfnamefont {J.~M.}\ \bibnamefont {Nichol}}, \bibinfo {author} {\bibfnamefont {S.~D.}\ \bibnamefont {Bartlett}}, \bibinfo {author} {\bibfnamefont {A.~C.}\ \bibnamefont {Doherty}}, \bibinfo {author} {\bibfnamefont {V.}~\bibnamefont {Umansky}},\ and\ \bibinfo {author} {\bibfnamefont {A.}~\bibnamefont {Yacoby}},\ }\bibfield  {title} {\bibinfo {title} {Suppressing qubit dephasing using real-time hamiltonian estimation},\ }\href {https://doi.org/10.1038/ncomms6156} {\bibfield  {journal} {\bibinfo  {journal} {Nature Communications}\ }\textbf {\bibinfo {volume} {5}},\ \bibinfo {pages} {5156} (\bibinfo {year} {2014})}\BibitemShut {NoStop}%
\bibitem [{\citenamefont {Zhai}\ \emph {et~al.}(2020)\citenamefont {Zhai}, \citenamefont {L\"{o}bl}, \citenamefont {Nguyen}, \citenamefont {Ritzmann}, \citenamefont {Javadi}, \citenamefont {Spinnler}, \citenamefont {Wieck}, \citenamefont {Ludwig},\ and\ \citenamefont {Warburton}}]{Zhai2020}%
  \BibitemOpen
  \bibfield  {author} {\bibinfo {author} {\bibfnamefont {L.}~\bibnamefont {Zhai}}, \bibinfo {author} {\bibfnamefont {M.~C.}\ \bibnamefont {L\"{o}bl}}, \bibinfo {author} {\bibfnamefont {G.~N.}\ \bibnamefont {Nguyen}}, \bibinfo {author} {\bibfnamefont {J.}~\bibnamefont {Ritzmann}}, \bibinfo {author} {\bibfnamefont {A.}~\bibnamefont {Javadi}}, \bibinfo {author} {\bibfnamefont {C.}~\bibnamefont {Spinnler}}, \bibinfo {author} {\bibfnamefont {A.~D.}\ \bibnamefont {Wieck}}, \bibinfo {author} {\bibfnamefont {A.}~\bibnamefont {Ludwig}},\ and\ \bibinfo {author} {\bibfnamefont {R.~J.}\ \bibnamefont {Warburton}},\ }\bibfield  {title} {\bibinfo {title} {Low-noise {GaAs} quantum dots for quantum photonics},\ }\href {https://doi.org/10.1038/s41467-020-18625-z} {\bibfield  {journal} {\bibinfo  {journal} {Nat. Commun.}\ }\textbf {\bibinfo {volume} {11}},\ \bibinfo {pages} {4745} (\bibinfo {year} {2020})}\BibitemShut {NoStop}%
\bibitem [{\citenamefont {K\"{u}ster}\ \emph {et~al.}(2016)\citenamefont {K\"{u}ster}, \citenamefont {Heyn}, \citenamefont {Ungeheuer}, \citenamefont {Juska}, \citenamefont {Tommaso~Moroni}, \citenamefont {Pelucchi},\ and\ \citenamefont {Hansen}}]{Kuster2016}%
  \BibitemOpen
  \bibfield  {author} {\bibinfo {author} {\bibfnamefont {A.}~\bibnamefont {K\"{u}ster}}, \bibinfo {author} {\bibfnamefont {C.}~\bibnamefont {Heyn}}, \bibinfo {author} {\bibfnamefont {A.}~\bibnamefont {Ungeheuer}}, \bibinfo {author} {\bibfnamefont {G.}~\bibnamefont {Juska}}, \bibinfo {author} {\bibfnamefont {S.}~\bibnamefont {Tommaso~Moroni}}, \bibinfo {author} {\bibfnamefont {E.}~\bibnamefont {Pelucchi}},\ and\ \bibinfo {author} {\bibfnamefont {W.}~\bibnamefont {Hansen}},\ }\bibfield  {title} {\bibinfo {title} {Droplet etching of deep nanoholes for filling with self-aligned complex quantum structures},\ }\href {https://doi.org/10.1186/s11671-016-1495-5} {\bibfield  {journal} {\bibinfo  {journal} {Nanoscale Res. Lett.}\ }\textbf {\bibinfo {volume} {11}},\ \bibinfo {pages} {282} (\bibinfo {year} {2016})}\BibitemShut {NoStop}%
\bibitem [{\citenamefont {Reindl}\ \emph {et~al.}(2019)\citenamefont {Reindl}, \citenamefont {Weber}, \citenamefont {Huber}, \citenamefont {Schimpf}, \citenamefont {Covre~da Silva}, \citenamefont {Portalupi}, \citenamefont {Trotta}, \citenamefont {Michler},\ and\ \citenamefont {Rastelli}}]{Reindl2019}%
  \BibitemOpen
  \bibfield  {author} {\bibinfo {author} {\bibfnamefont {M.}~\bibnamefont {Reindl}}, \bibinfo {author} {\bibfnamefont {J.~H.}\ \bibnamefont {Weber}}, \bibinfo {author} {\bibfnamefont {D.}~\bibnamefont {Huber}}, \bibinfo {author} {\bibfnamefont {C.}~\bibnamefont {Schimpf}}, \bibinfo {author} {\bibfnamefont {S.~F.}\ \bibnamefont {Covre~da Silva}}, \bibinfo {author} {\bibfnamefont {S.~L.}\ \bibnamefont {Portalupi}}, \bibinfo {author} {\bibfnamefont {R.}~\bibnamefont {Trotta}}, \bibinfo {author} {\bibfnamefont {P.}~\bibnamefont {Michler}},\ and\ \bibinfo {author} {\bibfnamefont {A.}~\bibnamefont {Rastelli}},\ }\bibfield  {title} {\bibinfo {title} {Highly indistinguishable single photons from incoherently excited quantum dots},\ }\href {https://doi.org/10.1103/PhysRevB.100.155420} {\bibfield  {journal} {\bibinfo  {journal} {Phys. Rev. B}\ }\textbf {\bibinfo {volume} {100}},\ \bibinfo {pages} {155420} (\bibinfo {year} {2019})}\BibitemShut {NoStop}%
\bibitem [{\citenamefont {Lehner}\ \emph {et~al.}(2023)\citenamefont {Lehner}, \citenamefont {Seidelmann}, \citenamefont {Undeutsch}, \citenamefont {Schimpf}, \citenamefont {Manna}, \citenamefont {Gawełczyk}, \citenamefont {Covre~da Silva}, \citenamefont {Yuan}, \citenamefont {Stroj}, \citenamefont {Reiter}, \citenamefont {Axt},\ and\ \citenamefont {Rastelli}}]{Lehner2023}%
  \BibitemOpen
  \bibfield  {author} {\bibinfo {author} {\bibfnamefont {B.~U.}\ \bibnamefont {Lehner}}, \bibinfo {author} {\bibfnamefont {T.}~\bibnamefont {Seidelmann}}, \bibinfo {author} {\bibfnamefont {G.}~\bibnamefont {Undeutsch}}, \bibinfo {author} {\bibfnamefont {C.}~\bibnamefont {Schimpf}}, \bibinfo {author} {\bibfnamefont {S.}~\bibnamefont {Manna}}, \bibinfo {author} {\bibfnamefont {M.}~\bibnamefont {Gawełczyk}}, \bibinfo {author} {\bibfnamefont {S.~F.}\ \bibnamefont {Covre~da Silva}}, \bibinfo {author} {\bibfnamefont {X.}~\bibnamefont {Yuan}}, \bibinfo {author} {\bibfnamefont {S.}~\bibnamefont {Stroj}}, \bibinfo {author} {\bibfnamefont {D.~E.}\ \bibnamefont {Reiter}}, \bibinfo {author} {\bibfnamefont {V.~M.}\ \bibnamefont {Axt}},\ and\ \bibinfo {author} {\bibfnamefont {A.}~\bibnamefont {Rastelli}},\ }\bibfield  {title} {\bibinfo {title} {Beyond the four-level model: Dark and hot states in quantum dots degrade photonic entanglement},\ }\href {https://doi.org/10.1021/acs.nanolett.2c04734} {\bibfield  {journal}
  {\bibinfo  {journal} {Nano Lett.}\ }\textbf {\bibinfo {volume} {23}},\ \bibinfo {pages} {1409} (\bibinfo {year} {2023})}\BibitemShut {NoStop}%
\bibitem [{\citenamefont {Dusanowski}\ \emph {et~al.}(2018)\citenamefont {Dusanowski}, \citenamefont {Gawe{\l}czyk}, \citenamefont {Misiewicz}, \citenamefont {H\"{o}fling}, \citenamefont {Reithmaier},\ and\ \citenamefont {S\k{e}k}}]{Dusanowski2018}%
  \BibitemOpen
  \bibfield  {author} {\bibinfo {author} {\bibfnamefont {{\L}.}~\bibnamefont {Dusanowski}}, \bibinfo {author} {\bibfnamefont {M.}~\bibnamefont {Gawe{\l}czyk}}, \bibinfo {author} {\bibfnamefont {J.}~\bibnamefont {Misiewicz}}, \bibinfo {author} {\bibfnamefont {S.}~\bibnamefont {H\"{o}fling}}, \bibinfo {author} {\bibfnamefont {J.~P.}\ \bibnamefont {Reithmaier}},\ and\ \bibinfo {author} {\bibfnamefont {G.}~\bibnamefont {S\k{e}k}},\ }\bibfield  {title} {\bibinfo {title} {Strongly temperature-dependent recombination kinetics of a negatively charged exciton in asymmetric quantum dots at 1.55 \textmu{}m},\ }\href {https://doi.org/10.1063/1.5030895} {\bibfield  {journal} {\bibinfo  {journal} {Appl. Phys. Lett.}\ }\textbf {\bibinfo {volume} {113}},\ \bibinfo {pages} {043103} (\bibinfo {year} {2018})}\BibitemShut {NoStop}%
\bibitem [{\citenamefont {Krenner}\ \emph {et~al.}(2005)\citenamefont {Krenner}, \citenamefont {Sabathil}, \citenamefont {Clark}, \citenamefont {Kress}, \citenamefont {Schuh}, \citenamefont {Bichler}, \citenamefont {Abstreiter},\ and\ \citenamefont {Finley}}]{Krenner2005}%
  \BibitemOpen
  \bibfield  {author} {\bibinfo {author} {\bibfnamefont {H.~J.}\ \bibnamefont {Krenner}}, \bibinfo {author} {\bibfnamefont {M.}~\bibnamefont {Sabathil}}, \bibinfo {author} {\bibfnamefont {E.~C.}\ \bibnamefont {Clark}}, \bibinfo {author} {\bibfnamefont {A.}~\bibnamefont {Kress}}, \bibinfo {author} {\bibfnamefont {D.}~\bibnamefont {Schuh}}, \bibinfo {author} {\bibfnamefont {M.}~\bibnamefont {Bichler}}, \bibinfo {author} {\bibfnamefont {G.}~\bibnamefont {Abstreiter}},\ and\ \bibinfo {author} {\bibfnamefont {J.~J.}\ \bibnamefont {Finley}},\ }\bibfield  {title} {\bibinfo {title} {Direct observation of controlled coupling in an individual quantum dot molecule},\ }\href {https://doi.org/10.1103/PhysRevLett.94.057402} {\bibfield  {journal} {\bibinfo  {journal} {Phys. Rev. Lett.}\ }\textbf {\bibinfo {volume} {94}},\ \bibinfo {pages} {057402} (\bibinfo {year} {2005})}\BibitemShut {NoStop}%
\bibitem [{\citenamefont {Bopp}\ \emph {et~al.}(2023)\citenamefont {Bopp}, \citenamefont {Schall}, \citenamefont {Bart}, \citenamefont {V\"ogl}, \citenamefont {Cullip}, \citenamefont {Sbresny}, \citenamefont {Boos}, \citenamefont {Thalacker}, \citenamefont {Lienhart}, \citenamefont {Rodt}, \citenamefont {Reuter}, \citenamefont {Ludwig}, \citenamefont {Wieck}, \citenamefont {Reitzenstein}, \citenamefont {M\"uller},\ and\ \citenamefont {Finley}}]{Bopp2023}%
  \BibitemOpen
  \bibfield  {author} {\bibinfo {author} {\bibfnamefont {F.}~\bibnamefont {Bopp}}, \bibinfo {author} {\bibfnamefont {J.}~\bibnamefont {Schall}}, \bibinfo {author} {\bibfnamefont {N.}~\bibnamefont {Bart}}, \bibinfo {author} {\bibfnamefont {F.}~\bibnamefont {V\"ogl}}, \bibinfo {author} {\bibfnamefont {C.}~\bibnamefont {Cullip}}, \bibinfo {author} {\bibfnamefont {F.}~\bibnamefont {Sbresny}}, \bibinfo {author} {\bibfnamefont {K.}~\bibnamefont {Boos}}, \bibinfo {author} {\bibfnamefont {C.}~\bibnamefont {Thalacker}}, \bibinfo {author} {\bibfnamefont {M.}~\bibnamefont {Lienhart}}, \bibinfo {author} {\bibfnamefont {S.}~\bibnamefont {Rodt}}, \bibinfo {author} {\bibfnamefont {D.}~\bibnamefont {Reuter}}, \bibinfo {author} {\bibfnamefont {A.}~\bibnamefont {Ludwig}}, \bibinfo {author} {\bibfnamefont {A.~D.}\ \bibnamefont {Wieck}}, \bibinfo {author} {\bibfnamefont {S.}~\bibnamefont {Reitzenstein}}, \bibinfo {author} {\bibfnamefont {K.}~\bibnamefont {M\"uller}},\ and\ \bibinfo {author} {\bibfnamefont {J.~J.}\ \bibnamefont
  {Finley}},\ }\bibfield  {title} {\bibinfo {title} {Coherent driving of direct and indirect excitons in a quantum dot molecule},\ }\href {https://doi.org/10.1103/PhysRevB.107.165426} {\bibfield  {journal} {\bibinfo  {journal} {Phys. Rev. B}\ }\textbf {\bibinfo {volume} {107}},\ \bibinfo {pages} {165426} (\bibinfo {year} {2023})}\BibitemShut {NoStop}%
\bibitem [{\citenamefont {Heyn}\ \emph {et~al.}(2017)\citenamefont {Heyn}, \citenamefont {K\"uster}, \citenamefont {Ungeheuer}, \citenamefont {Gr\"afenstein},\ and\ \citenamefont {Hansen}}]{HeynPRB2017}%
  \BibitemOpen
  \bibfield  {author} {\bibinfo {author} {\bibfnamefont {C.}~\bibnamefont {Heyn}}, \bibinfo {author} {\bibfnamefont {A.}~\bibnamefont {K\"uster}}, \bibinfo {author} {\bibfnamefont {A.}~\bibnamefont {Ungeheuer}}, \bibinfo {author} {\bibfnamefont {A.}~\bibnamefont {Gr\"afenstein}},\ and\ \bibinfo {author} {\bibfnamefont {W.}~\bibnamefont {Hansen}},\ }\bibfield  {title} {\bibinfo {title} {Excited-state indirect excitons in {GaAs} quantum dot molecules},\ }\href {https://doi.org/10.1103/PhysRevB.96.085408} {\bibfield  {journal} {\bibinfo  {journal} {Phys. Rev. B}\ }\textbf {\bibinfo {volume} {96}},\ \bibinfo {pages} {085408} (\bibinfo {year} {2017})}\BibitemShut {NoStop}%
\bibitem [{\citenamefont {Lienhart}\ \emph {et~al.}(2025)\citenamefont {Lienhart}, \citenamefont {Gawarecki}, \citenamefont {Stöcker}, \citenamefont {Bopp}, \citenamefont {Cullip}, \citenamefont {Akhlaq}, \citenamefont {Thalacker}, \citenamefont {Schall}, \citenamefont {Rodt}, \citenamefont {Ludwig}, \citenamefont {Reuter}, \citenamefont {Reitzenstein}, \citenamefont {Müller}, \citenamefont {Machnikowski},\ and\ \citenamefont {Finley}}]{2505.09906}%
  \BibitemOpen
  \bibfield  {author} {\bibinfo {author} {\bibfnamefont {M.}~\bibnamefont {Lienhart}}, \bibinfo {author} {\bibfnamefont {K.}~\bibnamefont {Gawarecki}}, \bibinfo {author} {\bibfnamefont {M.}~\bibnamefont {Stöcker}}, \bibinfo {author} {\bibfnamefont {F.}~\bibnamefont {Bopp}}, \bibinfo {author} {\bibfnamefont {C.}~\bibnamefont {Cullip}}, \bibinfo {author} {\bibfnamefont {N.}~\bibnamefont {Akhlaq}}, \bibinfo {author} {\bibfnamefont {C.}~\bibnamefont {Thalacker}}, \bibinfo {author} {\bibfnamefont {J.}~\bibnamefont {Schall}}, \bibinfo {author} {\bibfnamefont {S.}~\bibnamefont {Rodt}}, \bibinfo {author} {\bibfnamefont {A.}~\bibnamefont {Ludwig}}, \bibinfo {author} {\bibfnamefont {D.}~\bibnamefont {Reuter}}, \bibinfo {author} {\bibfnamefont {S.}~\bibnamefont {Reitzenstein}}, \bibinfo {author} {\bibfnamefont {K.}~\bibnamefont {Müller}}, \bibinfo {author} {\bibfnamefont {P.}~\bibnamefont {Machnikowski}},\ and\ \bibinfo {author} {\bibfnamefont {J.~J.}\ \bibnamefont {Finley}},\ }\href@noop {} {\bibinfo {title}
  {Resonant and anti-resonant exciton-phonon coupling in quantum dot molecules}} (\bibinfo {year} {2025}),\ \Eprint {https://arxiv.org/abs/arXiv:2505.09906} {arXiv:2505.09906} \BibitemShut {NoStop}%
\bibitem [{\citenamefont {Boyer de~la Giroday}\ \emph {et~al.}(2011)\citenamefont {Boyer de~la Giroday}, \citenamefont {Sk\"old}, \citenamefont {Stevenson}, \citenamefont {Farrer}, \citenamefont {Ritchie},\ and\ \citenamefont {Shields}}]{BoyerPRL2011}%
  \BibitemOpen
  \bibfield  {author} {\bibinfo {author} {\bibfnamefont {A.}~\bibnamefont {Boyer de~la Giroday}}, \bibinfo {author} {\bibfnamefont {N.}~\bibnamefont {Sk\"old}}, \bibinfo {author} {\bibfnamefont {R.~M.}\ \bibnamefont {Stevenson}}, \bibinfo {author} {\bibfnamefont {I.}~\bibnamefont {Farrer}}, \bibinfo {author} {\bibfnamefont {D.~A.}\ \bibnamefont {Ritchie}},\ and\ \bibinfo {author} {\bibfnamefont {A.~J.}\ \bibnamefont {Shields}},\ }\bibfield  {title} {\bibinfo {title} {Exciton-spin memory with a semiconductor quantum dot molecule},\ }\href {https://doi.org/10.1103/PhysRevLett.106.216802} {\bibfield  {journal} {\bibinfo  {journal} {Phys. Rev. Lett.}\ }\textbf {\bibinfo {volume} {106}},\ \bibinfo {pages} {216802} (\bibinfo {year} {2011})}\BibitemShut {NoStop}%
\bibitem [{\citenamefont {Haken}(1975)}]{Haken1975}%
  \BibitemOpen
  \bibfield  {author} {\bibinfo {author} {\bibfnamefont {H.}~\bibnamefont {Haken}},\ }\bibfield  {title} {\bibinfo {title} {Statistical physics of bifurcation, spatial structures, and fluctuations of chemical reactions},\ }\href {https://doi.org/10.1007/bf01313213} {\bibfield  {journal} {\bibinfo  {journal} {Z. Phys. B}\ }\textbf {\bibinfo {volume} {20}},\ \bibinfo {pages} {413–420} (\bibinfo {year} {1975})}\BibitemShut {NoStop}%
\bibitem [{\citenamefont {Cummins}\ \emph {et~al.}(2003)\citenamefont {Cummins}, \citenamefont {Llewellyn},\ and\ \citenamefont {Jones}}]{CumminsPRA2003}%
  \BibitemOpen
  \bibfield  {author} {\bibinfo {author} {\bibfnamefont {H.~K.}\ \bibnamefont {Cummins}}, \bibinfo {author} {\bibfnamefont {G.}~\bibnamefont {Llewellyn}},\ and\ \bibinfo {author} {\bibfnamefont {J.~A.}\ \bibnamefont {Jones}},\ }\bibfield  {title} {\bibinfo {title} {Tackling systematic errors in quantum logic gates with composite rotations},\ }\href {https://doi.org/10.1103/PhysRevA.67.042308} {\bibfield  {journal} {\bibinfo  {journal} {Phys. Rev. A}\ }\textbf {\bibinfo {volume} {67}},\ \bibinfo {pages} {042308} (\bibinfo {year} {2003})}\BibitemShut {NoStop}%
\bibitem [{\citenamefont {Barnes}\ \emph {et~al.}(2015)\citenamefont {Barnes}, \citenamefont {Wang},\ and\ \citenamefont {Das~Sarma}}]{Barnes2015}%
  \BibitemOpen
  \bibfield  {author} {\bibinfo {author} {\bibfnamefont {E.}~\bibnamefont {Barnes}}, \bibinfo {author} {\bibfnamefont {X.}~\bibnamefont {Wang}},\ and\ \bibinfo {author} {\bibfnamefont {S.}~\bibnamefont {Das~Sarma}},\ }\bibfield  {title} {\bibinfo {title} {Robust quantum control using smooth pulses and topological winding},\ }\href {https://doi.org/10.1038/srep12685} {\bibfield  {journal} {\bibinfo  {journal} {Sci. Rep.}\ }\textbf {\bibinfo {volume} {5}},\ \bibinfo {pages} {12685} (\bibinfo {year} {2015})}\BibitemShut {NoStop}%
\bibitem [{\citenamefont {Khodjasteh}\ and\ \citenamefont {Viola}(2009)}]{Khodjasteh2009}%
  \BibitemOpen
  \bibfield  {author} {\bibinfo {author} {\bibfnamefont {K.}~\bibnamefont {Khodjasteh}}\ and\ \bibinfo {author} {\bibfnamefont {L.}~\bibnamefont {Viola}},\ }\bibfield  {title} {\bibinfo {title} {Dynamically error-corrected gates for universal quantum computation},\ }\href {https://doi.org/10.1103/PhysRevLett.102.080501} {\bibfield  {journal} {\bibinfo  {journal} {Phys. Rev. Lett.}\ }\textbf {\bibinfo {volume} {102}},\ \bibinfo {pages} {080501} (\bibinfo {year} {2009})}\BibitemShut {NoStop}%
\bibitem [{\citenamefont {Green}\ \emph {et~al.}(2013)\citenamefont {Green}, \citenamefont {Sastrawan}, \citenamefont {Uys},\ and\ \citenamefont {Biercuk}}]{Green2013}%
  \BibitemOpen
  \bibfield  {author} {\bibinfo {author} {\bibfnamefont {T.~J.}\ \bibnamefont {Green}}, \bibinfo {author} {\bibfnamefont {J.}~\bibnamefont {Sastrawan}}, \bibinfo {author} {\bibfnamefont {H.}~\bibnamefont {Uys}},\ and\ \bibinfo {author} {\bibfnamefont {M.~J.}\ \bibnamefont {Biercuk}},\ }\bibfield  {title} {\bibinfo {title} {Arbitrary quantum control of qubits in the presence of universal noise},\ }\href {https://doi.org/10.1088/1367-2630/15/9/095004} {\bibfield  {journal} {\bibinfo  {journal} {New J. Phys.}\ }\textbf {\bibinfo {volume} {15}},\ \bibinfo {pages} {095004} (\bibinfo {year} {2013})}\BibitemShut {NoStop}%
\end{thebibliography}%

\onecolumngrid

\section*{End Matter}
\renewcommand{\theequation}{A\arabic{equation}}
\setcounter{equation}{0}
{\label{sec:B}\it Appendix A: Shape of acoustic pulses---}The envelopes of flat-top pulses used in numerical simulations are given by
\begin{equation}
    f_i(t) = \frac{1}{2}\left[1-\erf\left({\frac{t - t_i' - \Delta t_i/2}{\kappa}}\right)\erf\left({\frac{t - t_i' + \Delta t_i/2}{\kappa}}\right)\right],
\end{equation}
where 
$t_i'$ is the $i$th pulse center, $\kappa$ the switching rate, and $\Delta t_i$ the pulse duration. For series of $M$ pulses $t_i'=\sum_{j=1}^{i} \Delta t_j - (\Delta t_1+\Delta t_i)/2$, and the total envelope is $f(t) = \sum_{i=1}^{M} f_i(t)$.

\renewcommand{\theequation}{B\arabic{equation}}
\setcounter{equation}{0}
{\label{sec:A}\it Appendix B: Adiabatic elimination---}We begin from writing the Schr{\"o}dinger equation for the system
$i\hbar\dot{\bm{\psi}}=\widetilde{H}_{\mathrm{c}}\bm{\psi}$ with $\bm{\psi}=(c_{\rightarrow}(t),c_{\leftarrow}(t),c_{\mathrm{T}}(t))^{\top\!}$
and the Hamiltonian \eqref{eq:secularHamiltonian} rewritten as $\widetilde{H}_{\mathrm{c}}=-\hbar\Delta\ketbra{\mathrm{T}}{\mathrm{T}} + \hbar\left(\eta_{\rightarrow}\ketbra{\rightarrow}{\mathrm{T}} + \eta_{\leftarrow}\ketbra{\leftarrow}{\mathrm{T}} + \hc\right)$, with $\eta_{\rightarrow}= A_{\mathrm{L}}J_1(\AcoustAmpl)e^{-i\varphi}/2$, $\eta_{\leftarrow}= A_{\mathrm{L}}J_0(\AcoustAmpl)/2$.
We deal with a set of three differential equations, the third of which reads
\begin{equation}\label{eq:adiab-elim}
i \dot{c}_{\mathrm{T}}(t) = -\Delta c_{\mathrm{T}}(t) + \eta_{\rightarrow}^{*} c_{\rightarrow}(t) + \eta_{\leftarrow}^{*} c_{\leftarrow}(t).
\end{equation}
In the adiabatic elimination \cite{Haken1975}, we separate the timescales of the fast but weak and unimportant oscillations of $c_{\mathrm{T}}$ due to the large detuning $\Delta$ and the slow evolution governed by the coupling to $\ket{\rightarrow}$ and $\ket{\leftarrow}$ states. As the trion is initially unoccupied, we may estimate $\abs{c_{\mathrm{T}}}$ to be as small as $\abs{\eta/\Delta}$ (let $\eta_{\rightarrow}\simeq\eta_{\leftarrow}=\eta$). Neglecting the fast ``own'' evolution of the trion as averaging to zero, and only keeping the sought-for second-order processes, the retained contribution to $\abs{\dot{c}_{\mathrm{T}}}$ is of the order of $\abs{\eta^2/\Delta}$. These estimates yield $\abs{\dot{c}_{\mathrm{T}}} \ll \abs{\Delta c_{\mathrm{T}}}$ given $\abs{\eta/\Delta} \ll 1$. Thus, we set $\dot{c}_{\mathrm{T}}=0$ on the left-hand side of Eq.~\eqref{eq:adiab-elim} and then solve algebraically for $c_{\mathrm{T}}$ as a combination of $c_{\rightarrow}$ and $c_{\rightarrow}$,
\begin{equation}\label{eq:adiab-elim-res}
    c_{\mathrm{T}}(t) = \frac{\eta_{\rightarrow}^{*}}{\Delta} c_{\rightarrow}(t) + \frac{\eta_{\leftarrow}^{*}}{\Delta} c_{\leftarrow}(t).
\end{equation}

We can arrive at the same result formally. Let us factor out the fast oscillation, $c_{\mathrm{T}}(t) = \exp(i\Delta t)\tilde{c}_{\mathrm{T}}(t)$ and define $f(t)=\eta_{\rightarrow}^{*} c_{\rightarrow}(t) + \eta_{\leftarrow}^{*} c_{\leftarrow}(t) $. Now, we have $i \dot{\tilde{c}}_{\mathrm{T}}(t) = \exp({-i\Delta t})f(t)$,
which we formally integrate $\tilde{c}_{\mathrm{T}}(t)=-i\int_0^t\mathrm{d}\tau e^{-i\Delta\tau} f(\tau)$. Integrating by parts and returning to the original amplitude $c_{\mathrm{T}}$, we get
\begin{align}
    {c}_{\mathrm{T}}(t) = \frac{1}{\Delta}\left[ f(t) - e^{i\Delta t}f(0) - \int_0^t \mathrm{d}\tau \,e^{i\Delta(t-\tau)}\frac{\mathrm{d}f(\tau)}{\mathrm{d}\tau} \right].
\end{align}
We continue integrating by parts, producing a series
\begin{align}\label{eq-adiab-elim-final}
    {c}_{\mathrm{T}}(t) = \sum_{n=1}^{\infty} i^{3n+1} \frac{1}{\Delta^n }\left[ f^{(n-1)}(t) - e^{i\Delta t}f^{(n-1)}(0) \right],
\end{align}
where $f^{(n)}(t)=\mathrm{d}^nf/\mathrm{d}t^n$. This form conveniently separates the adiabatic and oscillatory contributions. The consecutive derivatives $\abs{f^{(n-1)}}$ are of order $\abs{\eta}^{n}$, with $\eta$ representing the characteristic frequency of adiabatic evolution. We thus obtained a series expansion in the small parameter $\abs{\eta/\Delta}$, where adiabatic elimination retains the leading nonoscillatory term, reproducing \eqref{eq:adiab-elim-res}.

By making the trion state amplitude evolution fully dependent on the other two amplitudes, we reduce the problem to a Schr{\"o}dinger equation for an effective two-level system, $i\hbar(\dot{c}_{\rightarrow}(t),\dot{c}_{\leftarrow}(t))^{\top\!}=H_{\mathrm{spin}}(c_{\rightarrow}(t),c_{\leftarrow}(t))^{\top\!}$, with $H_{\mathrm{spin}}$ from \eqnref{eq:Hspin} describing a second-order acousto-optical coupling between spin states.

\renewcommand{\theequation}{C\arabic{equation}}
\setcounter{equation}{0}
{\label{sec:C}\it Appendix C: Inclination of rotation axis---}To extract the qubit rotation axis from the full numerical simulation, we identify the lowest point in the qubit trajectory (i.e., nearest to the south pole of the Bloch sphere, $\ket{\leftarrow}$ state) and pair it with the extremal point of the initial state, $\ket{\rightarrow}$. The vector normal to the line connecting these points defines the qubit rotation axis.

\begin{ruledtabular}
\begin{table}[tb!]
\centering
\small
\begin{tabular}{clllll}
& System & $T_2^*$ (ns) & $\hbar\sigma_{\delta}$ (neV) & Cooling method & Reference \\[2pt]
\hline
\symPurpleDiam & InAs/GaAs (self-assembled) & 39 & 23.9 & coherent population trapping feedback & \cite{EthierMajcherPRL2017} \\
\symGreenTri & InAs/GaAs (self-assembled) & $\sim$80$^\dagger$ & 11.6 &hole-assisted dynamical nuclear polarization & \cite{SunPRL2012} \\
\symBlueSquare & InAs/GaAs (self-assembled) & 296 & 3.14 &quantum-algorithmic feedback & \cite{JacksonPRX2022} \\
\symOrangeStar & GaAs/AlGaAs (droplet-etched) & 78 & 11.9 & Rabi cooling & \cite{NguyenPRL2023} \\
\symRedCircle & GaAs/AlGaAs (droplet-etched) & 608 & 1.53 & quantum-sensing-based cooling & \cite{NguyenPRL2023} \\
\end{tabular}
\caption{Electron-spin $T_2^*$ values and corresponding Overhauser root mean square energy in QD platforms. Symbols in the first column refer to labels in \subfigref{fig:fidelity}{a}. $^\dagger$For Ref.~\onlinecite{SunPRL2012}, $T_2^*$ is converted from the reported linewidth via $T_2^*\!=\!1/\gamma_s$ with $\gamma_s/2\pi=2~\text{MHz}$.\label{tab:app-T2star}\vspace{-0.5em}}
\end{table}
\end{ruledtabular}

\renewcommand{\theequation}{D\arabic{equation}}
\setcounter{equation}{0}
{\label{sec:D}\it Appendix D: Spin $T_2^{*}$ values---}In the simulations, we calculate the root mean square Overhauser field energy as $\hbar\sigma_{\delta} = \sqrt{2}\hbar/T_2^{*}$ based on the literature $T_2^{*}$ times (assuming hyperfine-limited) for InAs and GaAs QDs with cooled nuclear spins. Table~\ref{tab:app-T2star} lists these values, including their sources and cooling methods.

\begin{table}[tb!]
\caption{Pulse parameters, $\varphi_i$ and $\alpha_i$ and $\sigma_i$ for the $X$-gate sequence that cancels the leading-order longitudinal quasi-static Overhauser noise.\vspace{-0.5em}}
\label{tab:app-pulses}
\begin{ruledtabular}
\begin{tabular}{ccccccccc}
Pulse index $i$ & 1 & 2 & 3 & 4 & 5 & 6 & 7 \\
\hline
$\varphi_i/\pi$ & $1.0656$ & $0.9867$ & $0.8159$ & $0.3096$ & $1.9170$ & $1.6056$ & $0.6490$ \\
$\alpha_i/\pi$  & $1.0550$ & $0.5589$ & $0.1152$ & $0.6825$ & $0.3454$ & $0.5169$ & $0.5111$ \\
$\Delta t_i$~(ps)  & $12406.7$ & $6573.00$ & $1355.05$ & $8026.48$ & $4062.43$ & $6078.64$ & $6010.35$ \\
\end{tabular}
\end{ruledtabular}
\end{table}

\renewcommand{\theequation}{E\arabic{equation}}
\setcounter{equation}{0}
{\label{sec:E}\it Appendix E: Example longitudinal noise canceling sequence---} 
Standard CORPSE-like sequences \cite{CumminsPRA2003} do not apply to our tilted rotation axes, and typical methods require continuous tuning of both the inclination and azimuth \cite{Barnes2015,ZengPRA2019}. Here, we show that one can maintain a fixed inclination (set by the acoustic amplitude) while piecewise altering the azimuth through feasible phase skips on the acoustic drive.
The pulse sequence used here to implement the $X$ gate while canceling leading-order longitudinal quasi-static noise is ideally represented (for square pulses) as $R_{\varphi_N}(\alpha_N)\dots R_{\varphi_1}(\alpha_1)R_{\varphi_1}(\alpha_1)$ with $N=7$. Here, $\varphi$ defines the azimuth of the $i$th rotation axis (equal to the acoustic phase), always inclined by $\theta=\pi/4$, and $\alpha_i$ are the rotation angles. Table~\ref{tab:app-pulses} lists all $\varphi_i$ and $\alpha_i$ found by minimizing the leading-order contribution to gate infidelity (see below). The sequence was then simulated with realistic pulses with $\kappa=250$~ps. Pulse durations $\Delta t_i$ are calculated from $\alpha_i$ based on the duration of a $\pi$ rotation of $11.7599$~ns, with values in the last two rows of Table~\ref{tab:app-pulses}. No further optimization of the sequence was performed; only the laser parameters were optimized to compensate for the non-ideal acoustic pulses, yielding $\hbar\Delta \approx 0.76593$~meV, $\hbar A_{\mathrm{L}} \approx 37.9335$~\textmu{}eV.

\renewcommand{\theequation}{F\arabic{equation}}
\setcounter{equation}{0}
{\label{sec:F}\it Appendix F: Mitigating quasi-static noise impact---}
We adopt the approach from Refs.~\onlinecite{Khodjasteh2009,Green2013}. The control Hamiltonian for a single pulse inducing rotation about axis $\hat{\bm{n}}$ is $H_{\mathrm{c}}=(\hbar\Omega/2)\hat{\bm{n}}\cdot\bm{\sigma}$ and for duration $\alpha/\Omega$ induces evolution $U_{\mathrm{c}}(\alpha,\hat{\bm{n}})=\cos(\alpha/2)\mathbb{1}-i\sin(\alpha/2)\hat{\bm{n}}\cdot\bm{\sigma}$. The actual evolution during a pulse, driven by $H=H_{\mathrm{c}}+H_{\delta}$ is given by $U(t)=\exp(-i H t /\hbar)$ and differs from the ideal by the error propagator
\begin{equation}
    U_{\mathrm{err}}(t)=U_{\mathrm{c}}(t)^\dagger U(t).
\end{equation}
Differentiating shows that 
\begin{equation}
    \dot{U}_{\mathrm{err}}(t)=\left(-i/\hbar\right)\mathcal{H}(t)U_{\mathrm{err}}(t) \quad \text{with} \quad\mathcal{H}=U_{\mathrm{c}}(t)^\dagger H_{\delta} U_{\mathrm{c}}(t),
\end{equation}
i.e., the error evolution is driven by the noise Hamiltonian in the interaction picture of the control Hamiltonian (the toggling frame). For longitudinal noise, $H_{\delta}=(\hbar\delta/2)\hat{\bm{e}}_z\cdot\bm{\sigma}$, we have $\mathcal{H}(t)=(\hbar\delta/2)[R^{\top\!}(t)\hat{\bm{e}}_z]\cdot\bm{\sigma}$, where $R(t)$ is the Bloch-sphere rotation performed by $U_{\mathrm{c}}(t)$. Next, we express $U_{\mathrm{err}}(t)$ as a time-ordered exponential and apply the Magnus expansion,
\begin{equation}
    U_{\mathrm{err}}(t) = \mathcal{T}\exp\left(-i\int_0^t\mathrm{d}\tau\mathcal{H}(\tau)/\hbar\right)=\exp\left(-i\Phi_1-i\Phi_2+\dots\right),
\end{equation}
with $\mathcal{T}$ denoting time ordering. Retaining the first-order term with $\Phi_1=\int_0^t\mathrm{d}\tau\mathcal{H}(\tau)/\hbar$ yields $U_{\mathrm{err}}(t)\approx\exp(-i\delta\bm{g}\cdot\bm{\sigma}/(2\Omega))$, characterized by a dimensionless error vector $\bm{g}=\Omega\int_0^t\mathrm{d}\tau R^{\top\!}(\tau)\hat{\bm{e}}_z$.
For a square pulse, the angle $\phi=\Omega t$ varies from 0 to $\alpha$ over pulse duration $T$. We may calculate 
\begin{equation}
    \bm{g} = \Omega\int_0^T\mathrm{d}\tau R^{\top\!}(\tau)\hat{\bm{e}}_z = \int_0^\alpha\mathrm{d}\phi R_{\hat{\bm{n}}}\left(-\phi\right)\hat{\bm{e}}_z = \sin\alpha\hat{\bm{e}}_z-\left[\hat{\bm{n}}\times\hat{\bm{e}}_z\right](1-\cos\alpha)+\hat{\bm{n}}\left[\hat{\bm{n}}\cdot\hat{\bm{e}}_z\right](\alpha-\sin\alpha).
\end{equation}
The error accumulates over the sequence. The toggling frame differs for each pulse, and for the $k$th pulse, it is given by $R_{\mathrm{pre}}^{(k)}$, the total rotation from $k-1$ previous pulses. The total error for $M$ pulses is the sum of contributions rotated to a common frame 
\begin{equation}
    \bm{g}_{\mathrm{tot}}=\sum_{k=1}^{M} \left(R_{\mathrm{pre}}^{(k)}\right)^{\top\!} \bm{g}^{(k)}.
\end{equation}
The sequence has to both cancel $\bm{g}_{\mathrm{tot}}$ and realize the desired target evolution $U_{\mathrm{tgt}}$. The mismatch between the target and ideal evolution $U_0=U_{\mathrm{c}}(\alpha_M,\hat{\bm{n}}_M)\dots U_{\mathrm{c}}(\alpha_2,\hat{\bm{n}}_2)U_{\mathrm{c}}(\alpha_1,\hat{\bm{n}}_1)$ is $U_{\mathrm{tgt}}^{\dagger}U_0\approx\exp(-i\bm{v}\cdot\bm{\sigma}/2)$.
For a fixed rotation axis inclination, $\bm{g}_{\mathrm{tot}}$ and $\bm{v}$ are functions of $M$ azimuths $\varphi_i$ and $M$ rotation angles $\alpha_i$, which we find by minimizing both vectors (weighted). 
\end{document}